\documentclass[journal=jacsat,manuscript=article]{achemso}

\usepackage{chemformula} 
\usepackage[T1]{fontenc} 
\usepackage{subfigure}
\usepackage{float}
\usepackage{array}
\author{Hamid Rajabalipanah}
\affiliation {Applied Electromagnetic Labratoary, School of Electrical Engineering, Iran University of Science and Technology, Iran, Tehran.}
\author{Ali Abdolali}
\affiliation {Applied Electromagnetic Labratoary, School of Electrical Engineering, Iran University of Science and Technology, Iran, Tehran.}
\email{Abdolali@iust.ac.ir}
\author{Javad Shabanpour}
\affiliation {School of Electrical Engineering, Iran University of Science and Technology, Iran, Tehran.}
\author{Ali Momeni}
\affiliation {Applied Electromagnetic Labratoary, School of Electrical Engineering, Iran University of Science and Technology, Iran, Tehran.}
\author{Ahmad cheldavi}
\affiliation {School of Electrical Engineering, Iran University of Science and Technology, Iran, Tehran.}

\title
  {\textbf{Addition Theorem Revisiting for Phase/Amplitude-Encoded Metasurfaces: Asymmetric Spatial Power Dividers}}

\begin{document}


\begin{abstract}
Recent years have witnessed an extraordinary spurt in attention toward the wave manipulating strategies revealed by coding metasurfaces  as they build up a bridge between the physical and digital worlds. Newly, it has been shown that when two different coding patterns responsible for doing separate missions are added together based on the superposition theorem, the mixed coding pattern will perform both missions at the same time. In this paper, via a semi-analytical procedure, we demonstrate that such a theorem is not necessarily valid for all possible functionalities with considering phase-only coding distributions and ignoring the element pattern function. By revisiting the addition theorem, we introduce the concept of asymmetric spatial power divider (ASPD) with arbitrary power ratio levels in which modulating both amplitude and phase of the meta-atoms is inevitable to fully control the power intensity pattern of the metasurface. Numerical simulations illustrate that the proposed ASPD driven by proper coding sequences can directly generate a desired number of beams with pre-determined orientations and power budgets. The phase/amplitude-encoded Pancharatnam-Berry meta-atoms realize the required coding pattern in each case and good conformity between simulations and theoretical predictions verifies the presented formalism. This work exposes a new opportunity to implement spatial power dividers for various applications such as multiple-target radar systems, beamforming networks, and multiple-input multiple-output (MIMO) communication.
\end{abstract}

\section{Introduction}
~~One of the favorites of humankind is to guide the electromagnetic (EM) waves in their desired direction. According to the inability of natural materials to provide exotic wave-matter interactions, metamaterials have paved the way to control the EM waves in an unprecedented manner \cite{engheta2006metamaterials,holloway2012overview} . Metamaterials are artificial sub-wavelength metal/dielectric composites which arm a platform to realize various rich applications including, but not limited to, invisibility cloaks \cite{chu2018hybrid,zhu2013one,kim2018full} , negative refraction \cite{smith2004metamaterials,zhang2009negative} , illusion \cite{lai2009illusion,mach2018magnetic} , beam deflection \cite{barati2018experimental,pfeiffer2014efficient}  or epsilon near zero behaviors \cite{pollard2009optical,alu2007epsilon} . To overcome the metamaterials' disadvantages arising from fabrication complexities, high inherent losses, strong dispersion and bulky profiles, metasurfaces have emerged as 2D version of metamaterials to offer a promising groundwork to tailor diverse signatures related to the EM waves like amplitude\cite{liu2014broadband,rahmanzadeh2017analytical} , phase\cite{gao2015broadband,rouhi2018real} , polarization\cite{zhao2011manipulating,yin2015ultra} ,  and wave-vector \cite{liu2018unidirectional,achouri2015general} . The initial overwhelming interest in metasurfaces lies in abrupt and controllable change of wavefronts through a spatial inhomogeneity made by sub-wavelength scatterers, called meta-atoms, over an infinitesimally thin interface whereby we are able to mold wavefronts into shapes that can be designed at will\cite{momeni2018generalized} .

In an effort to manipulate the EM waves with more degree of freedom, a simple but yet powerful concept was emerged as “Coding metasurface” \cite{cui2014coding} . The coding representation of metasurfaces remarkably facilitates the design and provides great convenience owing to the digitalization of meta-atom geometry where it also provides a conceptual link between the physical and digital worlds\cite{cui2016information} . By purposefully distributing the coding particles ordered by a certain coding pattern over a 2D plane, coding metasurfaces open the door to many novel and programmable functional devices such as abnormal mirrors \cite{wan2016field,forouzmand2017real,diaz2017generalized,chalabi2017efficient,wong2014design} , low scattering surfaces \cite{rouhi2018real,moccia2017coding,rajabalipanah2018circular} , and information encryption interfaces \cite{momeni2018information} in a more simple and efficient way, where the conventional strategies fail to achieve satisfactory performances. Recently, it was shown that when two different coding patterns are added together via the superposition theorem, the mixed coding pattern will elaborately perform both functionalities of primary metasurfaces at the same time\cite{ma2016multi} . More recently, it has been also demonstrated that involving the complex codes aids to reach multifunctional meta-reflectors, where several missions of coding metasurfaces could be flexibly superposed together by means of the addition theorem\cite{wu2018addition} . Nevertheless, in these versions, the addition operation of two complex digital codes just captured the phase information of the coding particles and the amplitude data were ignored. In this work, with revisiting the addition theorem in coding metasurfaces, we will describe that such an assumption would bring irreparable consequences when dealing with the power intensity pattern of metasurfaces.  

Along the direction of evolution in coding metasurfaces, it's time to manipulate the power intensity pattern of the metasurfaces, an appealing functionality that has not been reported, yet. In this study, it is also shown that the conventional beam  splitters\cite{zhang2018metasurface,nayeri2013design,gomez2014optimization} , compulsorily generate two/multiple beams with unwanted power ratios dictated by their tilt angles. Accessing to asymmetric spatial power dividers (ASPD) with full/independent control over the power level carried by each beam, however, can give us a fabulous flexibility and privilege and is highy demanded in different practical applications, such as satellite communications, multiple-target radar systems, and multiple-input multiple-output (MIMO) communication \cite{zhao2018programmable,tang2018wireless} . Formerly, a few studies have contributed to address multibeam reflectarrays with controllable power of reflected beams. For instance, Nayeri et al. \cite{nayeri2013design} proposed a single-feed reflecarray with asymmetric beam directions and power levels. However, this work has accompanied with brute-force optimization procedure which results in a high computational cost. More recently, Zhang et al. \cite{zhang2018metasurface} represented that by changing the angle  of incidence, the amount of power distribution for each beam can be varied. Nevertheless, once the power ratios are determined, the direction of the reflected beams is forced to the user. Meanwhile, the presented design, which is restricted to only two beams, is based on numerical predictions and no semi-analytical framework supports that study. To the best of authors' knowledge, the precise control of power budgets of multiple beams with arbitrary orientations has not been reported, yet, and quite remains as a challenging task.

In this paper, we revisit the addition principle in coding metasurfaces to prevail the above-mentioned shortcomings, where both amplitude and phase of the coding particles are engineered to control the power intensity pattern of the proposed ASPD. Based on the superposition theorem, a general and straightforward semi-analytical method is presented to predict the exact power level of each radiated beam. It is demonstrated that the ASPD design with phase-only coding meta-atoms fails to achieve satisfactory results. Moreover, the quantization/discretization and element factor effects are incorporated in our design where the numerical simulations depict that benefited from 3-bit phase/amplitude-encoded C-shaped particles, we can accurately manipulate the power ratios without sacrificing the meta-atom amplitude and changing the tilt angles. To demonstrate the generality of the concept, several illustrative examples are studied. The simulation results have a very good agreement with the theoretical predictions and we believe that the proposed ASPD will open up new opportunities to power-controlling applications. 
\begin{figure*}[tb]
	\centering
	\includegraphics[width=.9\textwidth]{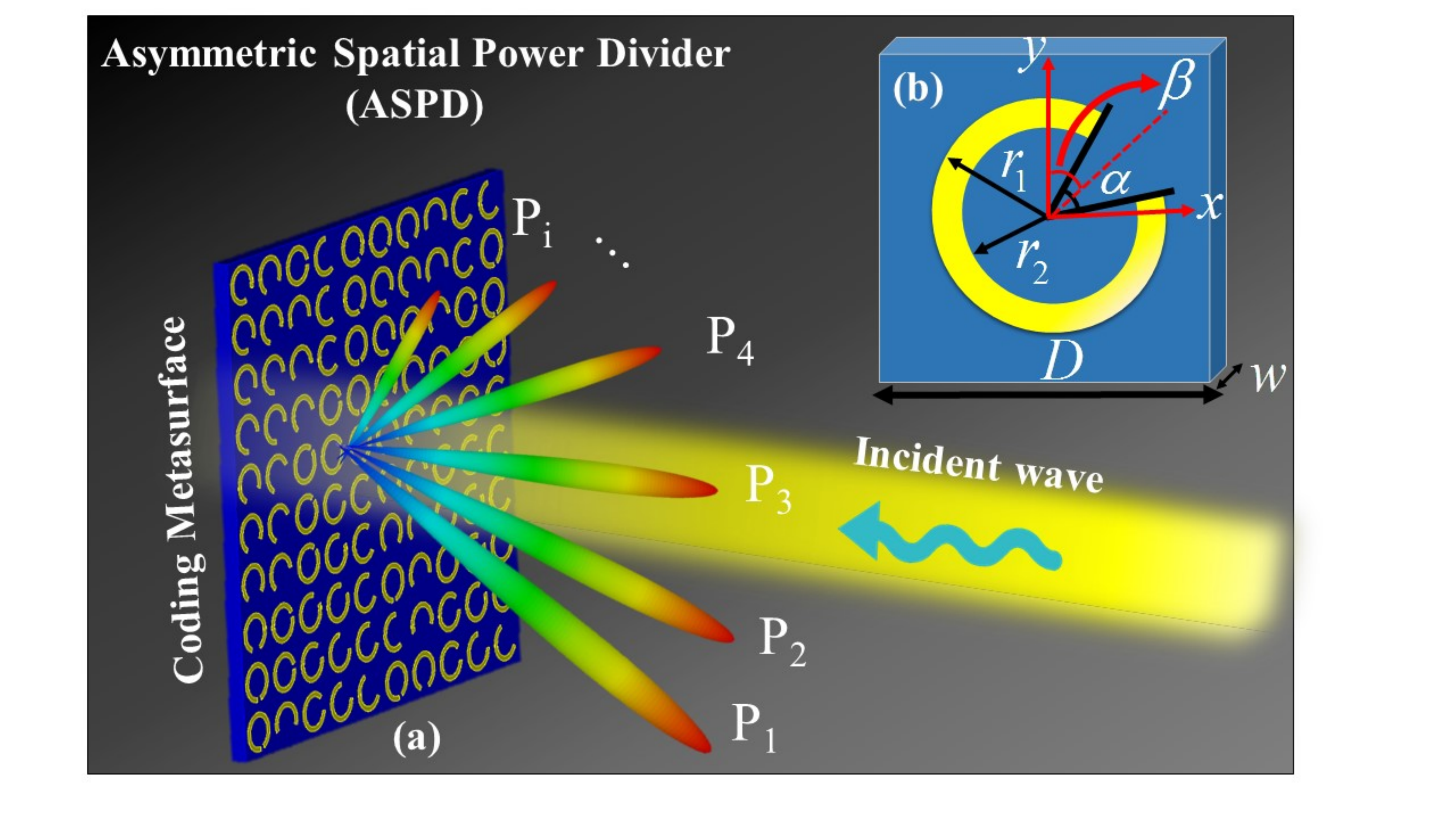}
	\caption{(a) The conceptual illustration of the proposed ASPD architecture  and (b) The front view of the employed meta-atom. The geometrical parameters are ${r_1}$ = 4mm, ${r_2} =$ 3mm, and $w =$ 3.2mm. }
	\label{Fig:Ts}
\end{figure*}
\section{Theoretical Framework}
~~Unlike the traditional “analog metamaterials” specified with continuous effective medium parameters, coding metasurfaces drastically facilitate the wave-matter interaction where the designs rely on introducing field discontinuity along the surface by spatially engineering the scattering meta-atoms in an array \cite{zhang2018digital} . By introducing abrupt phase shifts covering the range of $[0-2\pi]$, the coding metasurfaces with spatially varying geometries \cite{bao2018design} , DC biases \cite{momeni2018information} , and orientations \cite{zhang2017spin,liu2016anomalous} can imprint specified phase discontinuities on the propagating fields over the subwavelength scale, based on the generalized Snell’s law \cite{yu2011light} . Referring to this law, the in-plane component of the incident wave-vector is artificially mapped to that of the desired reflected one as $k_{r}^\parallel=(k_{0}\sin\theta \cos\phi+\nabla \phi_{x}) \hat{x}+(k_{0}\sin\theta \sin\phi+\nabla \phi_{y}) \hat{y}$ wherein the gradient symbols denote the slopes of the phase variations along the x and y directions, respectively. For $k_{r}^\parallel>k_{i}$, the anomalous reflection behavior appears in which the direction of the scattered beam can be determined by \cite{ding2017gradient}
\begin{align}
&\theta_r = \arcsin \,(\sqrt{k_{rx}^2+k_{ry}^2)}/{k_0})
&\phi_r = \arctan \,(k_{ry}/k_{rx})
\end{align}
\textbf{\textcolor{blue}{Fig. 1a}} demonstrates the conceptual illustration of the square metasurface generating multiple beams of different power levels that consists of $N \times N$  equal-sized meta-atoms with the periodicity of D along both vertical and horizontal directions. From the antenna array theory and upon illuminating by a normal plane wave, the far-field pattern function of the coding metasurface can be rigorously calculated as the superposition of the fields scattered by each contributing meta-atom \cite{jiu2017flexible}
 \begin{figure*}[t]
	$\begin{array}{rl}
	\includegraphics[width=1\textwidth]{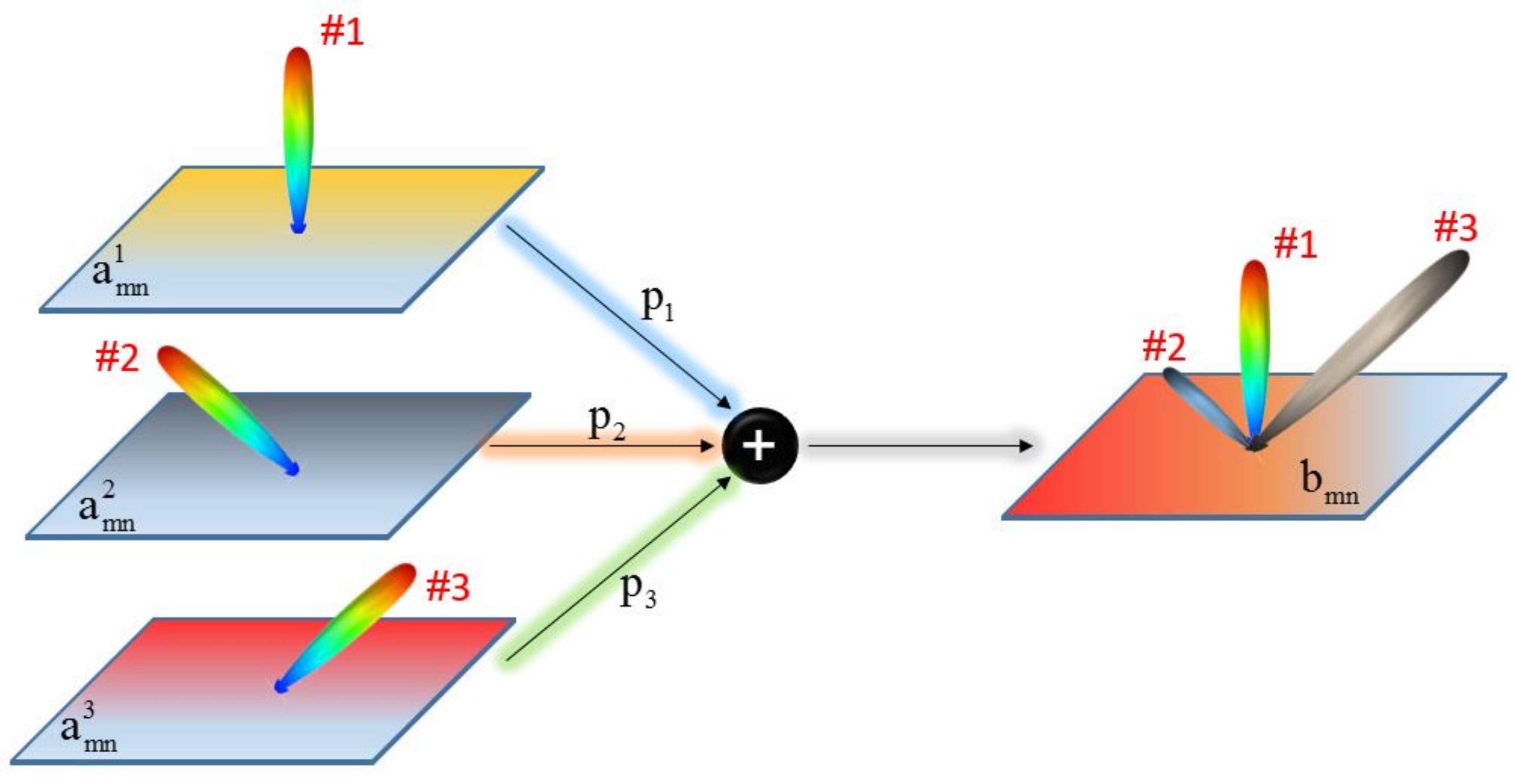} 
	
\end{array}$
\caption[.]{\label{fig:label} The schematic illustration of the addition theorem revisiting by incorporating the power coefficients to generate multiple asymmetric beams with arbitrary power ratio levels.}
\end{figure*}
\begin{equation}
{E_{scat}}(\theta ,\varphi ) = {E_{elem}}(\theta ,\varphi ) \times F(\theta ,\varphi )
\end{equation}
\begin{equation}
F(\theta ,\varphi ) = \sum\limits_{m = 1}^N {\sum\limits_{n = 1}^N {a_{mn}}\exp \{  - j\ (m - 1/2)~kD \sin \,\theta \cos \varphi \, - j \,(n - 1/2)~kD\,\sin \,\theta \sin \,\varphi \} \  } 
\end{equation}

In the above equation, $E_{element}(\theta,\phi)$ and ${a_{mn}}$ are, respectively, the pattern function and complex reflection coefficient of the meta-atoms, $F(\theta,\phi)$ refers to the array factor, $\theta$ and $\phi$ are the elevation and azimuth angles, and $k = 2\pi /\lambda $ is the wavenumber, where $\lambda $  is the working wavelength. The directivity function can be subsequently computed by \cite{liang2016broadband,wan2016field}

\begin{equation}
Dir(\theta ,\varphi ) = \frac{{4\pi {{\left| {F(\theta ,\varphi )} \right|}^2}}}{{\int\limits_0^{2\pi } {\int\limits_0^{\pi /2} {{{\left| {F(\theta ,\varphi )} \right|}^2}\,\sin \theta \,d\theta \,d\varphi } } }}
\end{equation}
~~~Regarding the Fourier connection between the coding pattern and its radiated beams  \cite{liu2016convolution} , many theorems in digital realm can be exploited for design of coding metasurfaces. For instance, based on the superposition of the aperture fields \cite{nayeri2013design,nayeri2012design} , the additive combination of different coding patterns, $a_{mn}^j$, yields a mixed coding pattern, $b_{mn}$, whereby all the individual missions will appear at the same time. To generate $M$ multiple functionalities governed by different coding patterns, simultaneously, the complex status of the constituent meta-atoms must read the following addition principle \cite{wu2018addition}
\begin{align}
& a_{mn}^{1}+a_{mn}^{2}+...+a_{mn}^{M}=b_{mn}^{{}}\\ 
&|a_{mn}^{1}|e^{j\phi_{mn}^{1}}+|a_{mn}^{2}|e^{j\phi_{mn}^{2}}+...+|a_{mn}^{M}|e^{j\phi_{mn}^{M}}=b_{mn}
\end{align}
~~~Here, $|a_{mn}^{j}|$ and ${\phi _{mn}^{j}}$  are the reflection amplitude and the reflection phase pertaining to the $(m,n)^{th}$ coding particle in $j^{th}$ coding pattern, and ${b_{mn}}$ denotes the complex digital status of coding particles in the mixed coding pattern. Without loss of generality, one can logically assume that the functionalities made by the primary coding patterns (like focusing, abnormal deflection and so on) do not require the amplitude varation of meta-atoms, i.e. $|a_{mn}^{j}|=1$ \cite{huang2018metasurface} . However, it will be shown that ignoring the amplitude information of $b_{mn}$ coefficients may encounter serious problems when we deal with the power intensity pattern of the coding metasurface. Hence, the mixed coding pattern must possess both phase and amplitude information of the combined coding pattern. By taking 2D IFFT from \textcolor{blue}{Eq. 6}
\begin{equation}
{IFFT \bigg( {{e}^{j\phi _{mn}^{1}}} \bigg)+IFFT\bigg( {{e}^{j\phi _{mn}^{2}}} \bigg)+...+IFFT\bigg( {{e}^{j\phi _{mn}^{M}}} \bigg)=IFFT\bigg( {{b}_{mn}} \bigg)} 
\end{equation}
Knowing that ${F^{j}}({\theta},{\varphi})\, = IFFT({e}^{j\phi _{mn}^{j}})$  ~\cite{momeni2018information} , then
\begin{equation}
{F^{1}}({\theta},{\varphi})\, + \,\,{F^{2}}({\theta},{\varphi})\, + \, \ldots \, + {F^{M}}({\theta},{\varphi})\, = \,{F_{sup}}({\theta},{\varphi})
\end{equation}
in which, $F_{sup}({\theta},{\varphi})$ stands for the superimposed array factor. Thus, the scattered fields of the combined coding metasurface can be written according to \textcolor{blue}{Eq. 2},
\begin{equation}
{E_{sup}(\theta,\phi)}=\sum\limits_{j = 1}^M {{E_{element}(\theta,\phi)}\,{F^{j}}({\theta},{\varphi})} 
\end{equation}

It should be noted that the multiple scattered beams created by the superimposed coding pattern obey the above equation. In the generalized revisiting of the addition principle, the power level of each radiating beam can be independently controlled via incorporating different multiplicative power constants, $\sqrt{p^{j}}$, to \textcolor{blue}{Eqs. 5 and 6}. A similar mathematical manipulation yields 
\begin{align}
& \sqrt{p^{1}}a_{mn}^{1}+\sqrt{p^{2}}a_{mn}^{2}+...+\sqrt{p^{M}}a_{mn}^{M}=b_{mn}^{{gen}}\\ 
&{E_{sup}^{gen}(\theta,\phi)}=\sum\limits_{j = 1}^M {\sqrt{p^{j}}{E_{element}(\theta,\phi)}\,{F^{j}}({\theta},{\varphi})} 
\end{align}
~~~The above equation represents the generalized form of addition theorem in which $E_{sup}^{gen}(\theta,\phi)$ caused by new combined coding pattern, $b_{mn}^{gen}$, contains $M$ multiple beams pointing at $(\theta^{j},\phi^{j})$ directions (see \textcolor{blue}{Fig. 2}). Subsequently, for each couple of pencil beams, the power ratio level obeys the following relation 
	 
\begin{equation}
\frac{P_{\sup }^{gen}\left( {{\theta }^{j}},{{\phi }^{j}} \right)}{P_{\sup }^{gen}\left( {{\theta }^{i}},{{\phi }^{i}} \right)}={{\bigg| \frac{E_{\sup }^{gen}\left( {{\theta }^{j}},{{\phi }^{j}} \right)}{E_{\sup }^{gen}\left( {{\theta }^{i}},{{\phi }^{i}} \right)} \bigg|}^{2}}=\frac{{{p}^{j}}}{{{p}^{i}}}\times {{\left| \frac{E_{element}^{{}}\left( {{\theta }^{j}},{{\phi }^{j}} \right)}{E_{element}^{{}}\left( {{\theta }^{i}},{{\phi }^{i}} \right)} \right|}^{2}} 
\end{equation}
~~~ where, $P_{\sup }^{gen}(\theta^{j},\phi^{j})$ and $P_{\sup }^{gen}(\theta^{i},\phi^{i})$   illustrate the peak power intensity of two arbitrarily-selected beams oriented along $(\theta^{j},\phi^{j})$ and $(\theta^{i},\phi^{i})$ directions, respectively. It should be noted that in deducing \textcolor{blue}{Eq. 12}, we assume that the angular position of the maximum in the scattering pattern of each coding pattern is located in the vicinity of the null of the other scattering patterns, i.e. $E_{scatt}^{j\ne{i}} \left( {{\theta^{i} }},{{\phi^{i} }} \right)\simeq 0$. It should be emphasized that \textcolor{blue}{Eq. 12} does not remain further valid if we ignore either the amplitude information of the mixed coding pattern, $|b_{mn}|$, or the pattern function of the meta-atoms. Nevertheless, this assumption was frequently utilized in the previous studies \cite{wu2018addition} .    
\section{ Concept Verification}
~~~~In particular, we assume the element pattern function as
 \begin{equation}
{{E}_{element}}\left( \theta ,\phi  \right)=\cos \left( \theta  \right)\times \sin c\bigg( \frac{1}{2}kd\sin \theta \cos \phi  \bigg)\times \sin c\bigg( \frac{1}{2}kd\sin \theta \sin \phi  \bigg)
 \end{equation}
~~Nevertheless, the cosine function of \textcolor{blue}{Eq. 13} closely resembles the element factor of the antenna elements \cite{moccia2017coding,zhang2018metasurface,momeni2018information} . With substituting  \textcolor {blue}{Eq. 13~} into \textcolor{blue}{Eq. 12}, one can readily deduce that in order to design an ASPD with desired power ratio level, the power coefficients should be chosen as
\begin{equation}
\frac{{{p^i}}}{{{p^j}}}\, = \,\sqrt {\frac{{{P_{sup}^{gen}(\theta^{i},\phi^{i})}}}{{{P_{sup}^{gen}(\theta^{j},\phi^{j})}}}}  \times \frac{{\cos \,({\theta^j})}}{{\cos \,({\theta^i})}}
\end{equation}
~~Once the power coefficients are determined, the phase and amplitude information of the superimposed coding pattern can be immediately obtained from \textcolor {blue}{Eq. 10}. Subsequently, the phase/amplitude-adjustable meta-atoms are aimed at imitating the required EM local responses dictated by the superimposed coding pattern. For the sake of simplicity, we put the focus of our design into the case of coding patterns with two and three asymmetric multiple beams to verify our claims about the necessity of the addition theorem revisiting. 
\begin{figure*}[t!]
	$\begin{array}{rl}
	\includegraphics[width=0.5\textwidth]{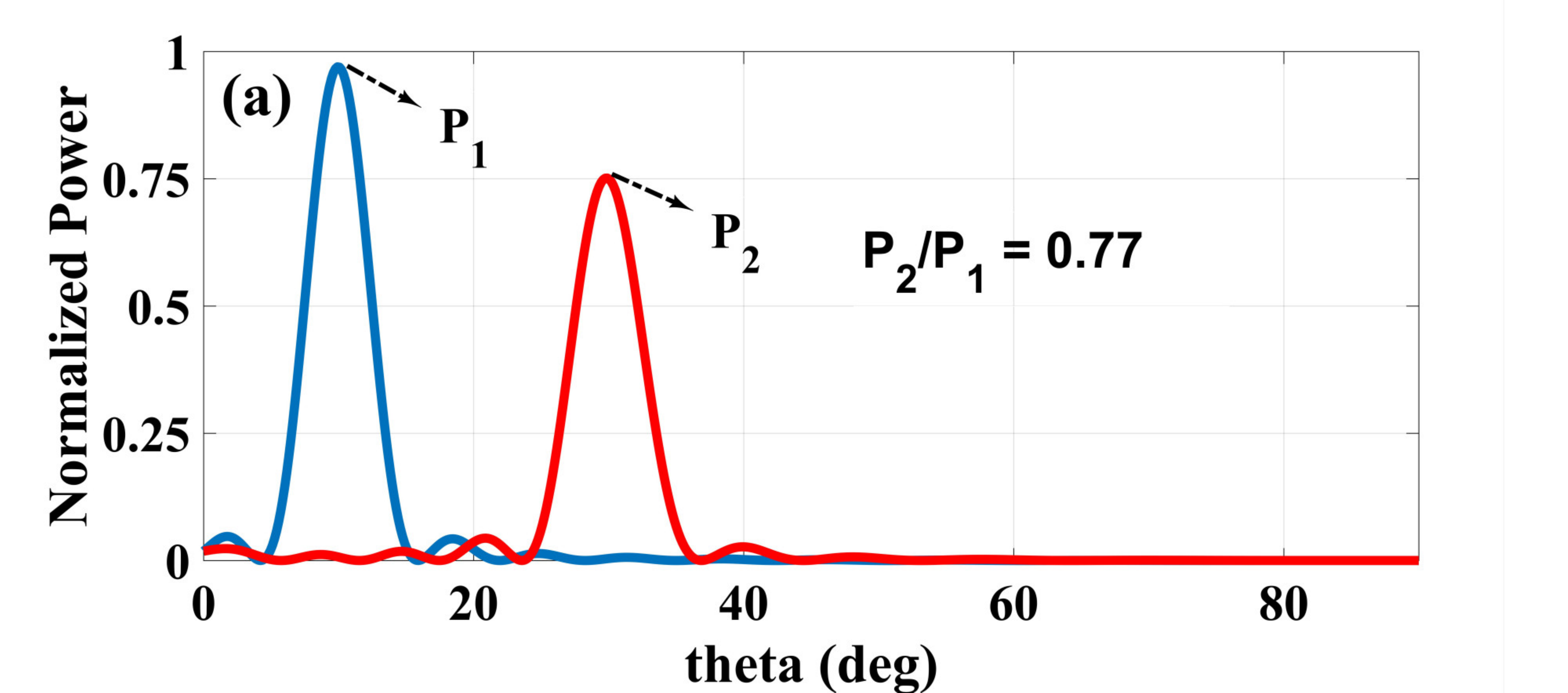} 
	\includegraphics[width=0.5\textwidth]{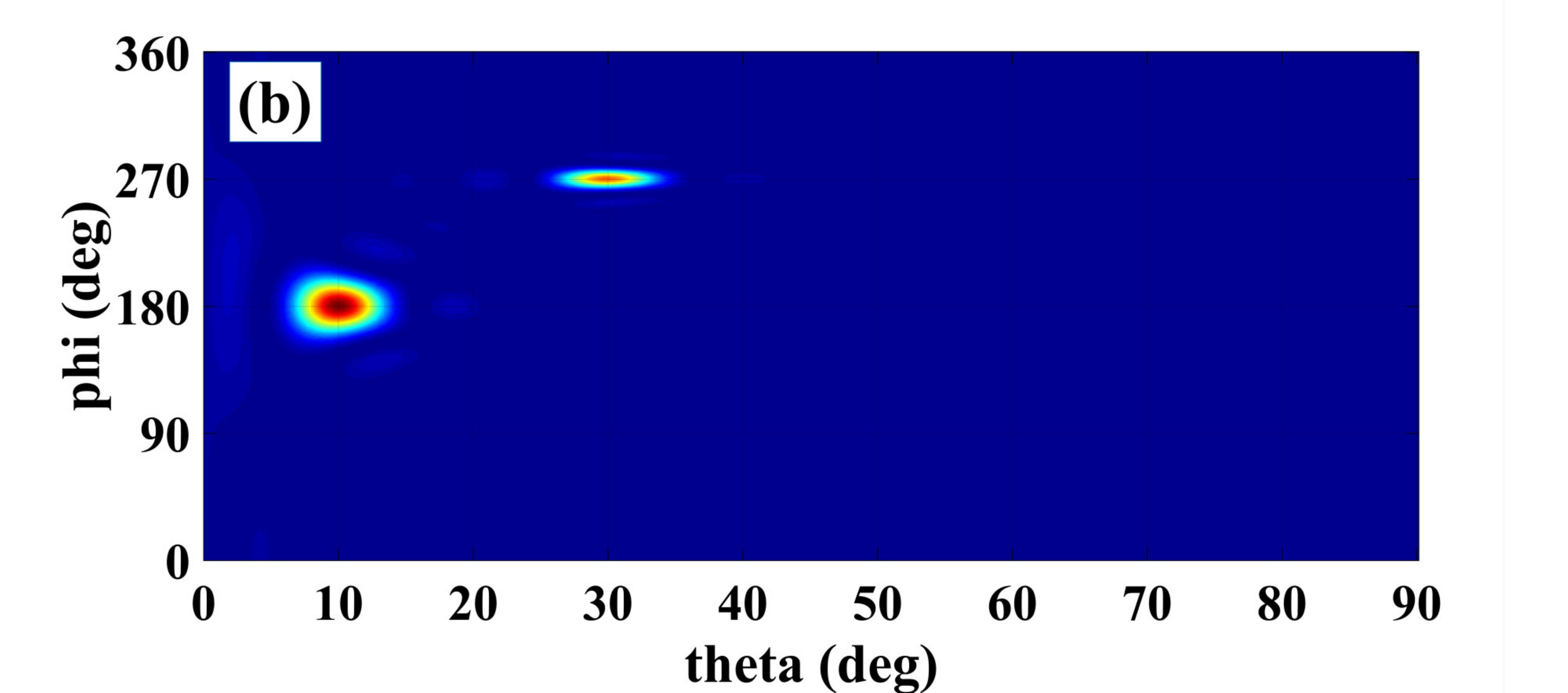} \\
	
	\includegraphics[width=0.5\textwidth]{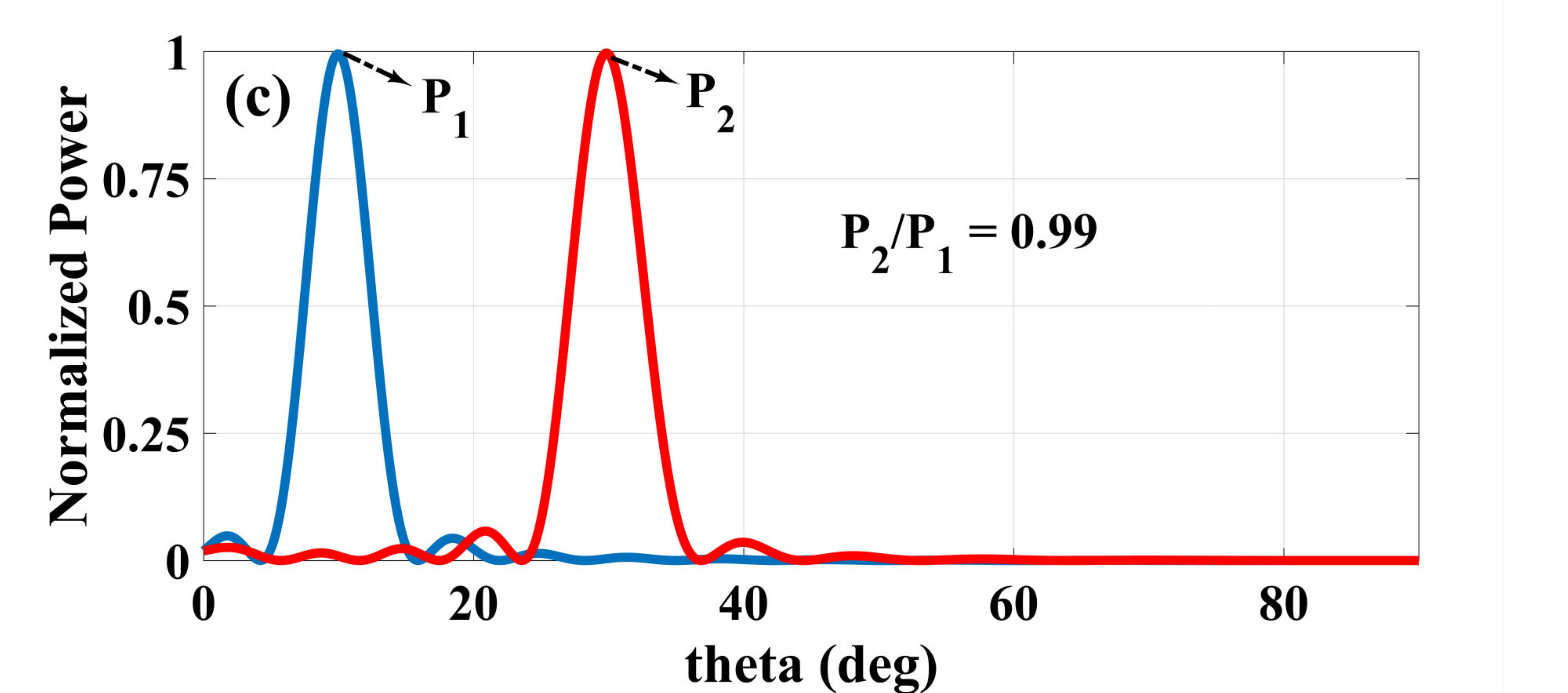}
	\includegraphics[width=0.5\textwidth]{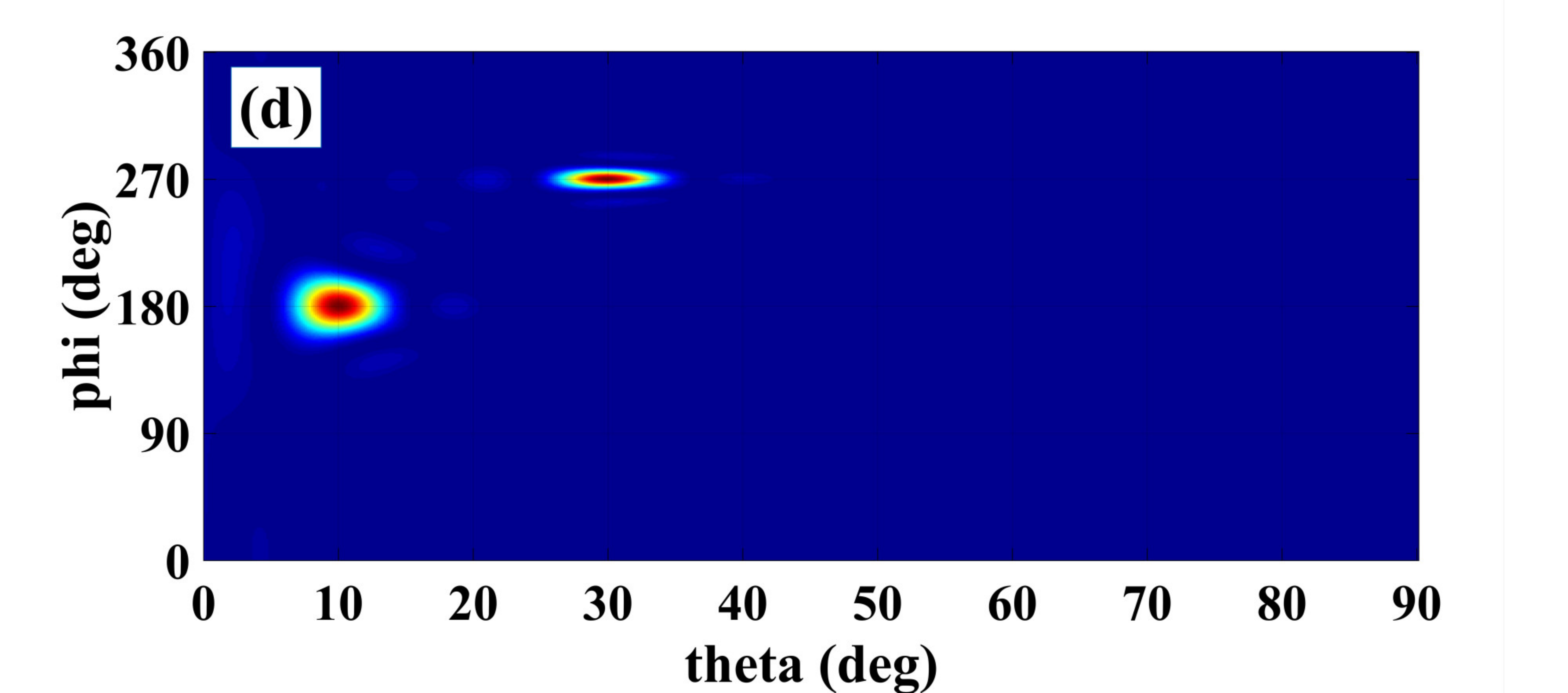} \\
	\includegraphics[width=0.5\textwidth]{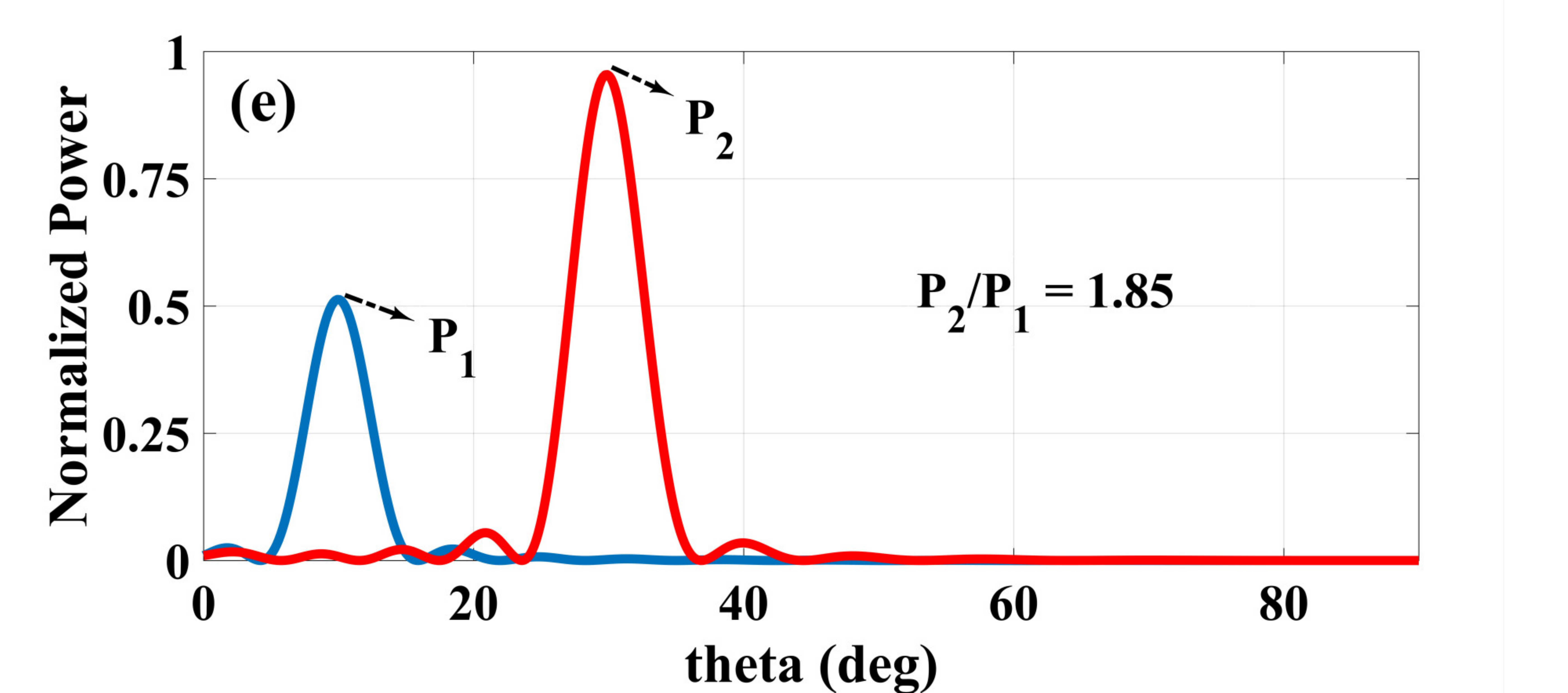}
	\includegraphics[width=0.5\textwidth]{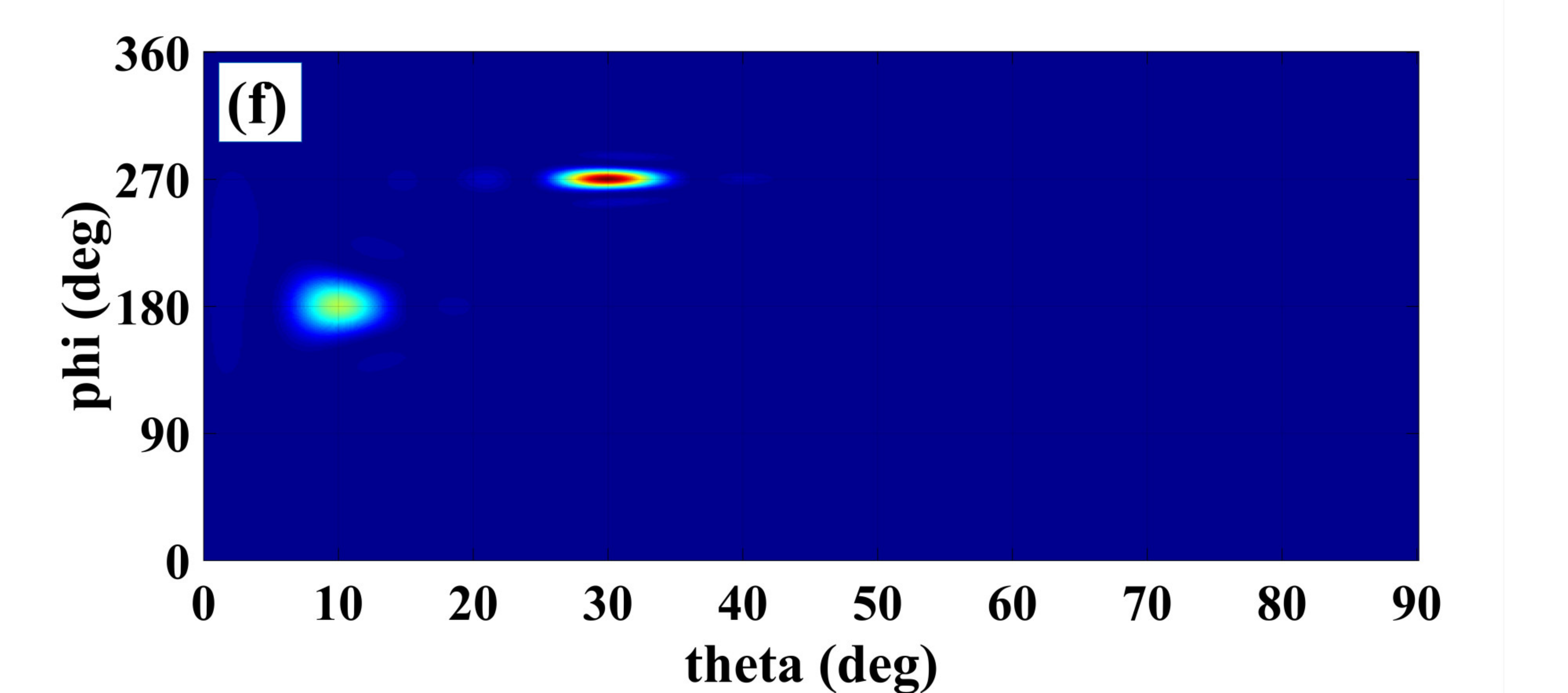}
\end{array}$
\caption[.]{\label{fig:label}Demonstration of three power-dividing examples for different power ratio levels via monitoring the 2D scattering patterns of the ASPD structure}
\end{figure*}

The numerical simulations are firstly carried out in the Matlab software using the antenna array theory. We begin the study with semi-continues designs in which the ASPD structures are imperceptibly (smoothly) discretized and no phase/amplitude quantization has been applied. The configuration of the semi-continues demonstrations is the same depicted in \textcolor{blue}{Fig. 2} in which all metasurfaces are composed of $N=200$ particles separated periodically at the distance of $D_{x}=D_{y}=\lambda/20$~~together (to approximately mimic a latterly-infinite continues phase/amplitude modulation). In the first illustration, we demonstrate how to achieve two asymmetrically oriented scattering beams with a specific power ratio by applying the generalized addition operation on the corresponding complex coding patterns.
\begin{figure*}[t!]
	$\begin{array}{rl}
	\includegraphics[width=0.5\textwidth]{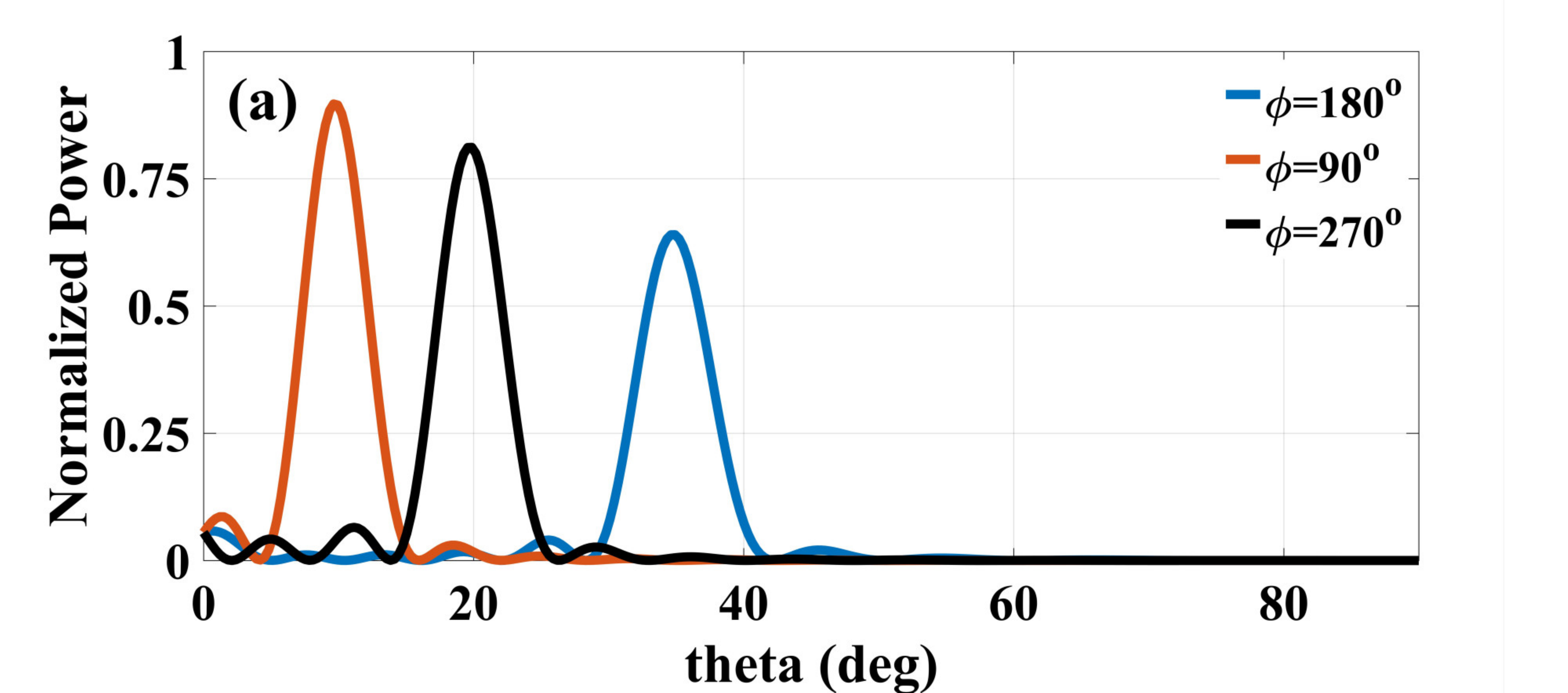} 
	\includegraphics[width=0.5\textwidth]{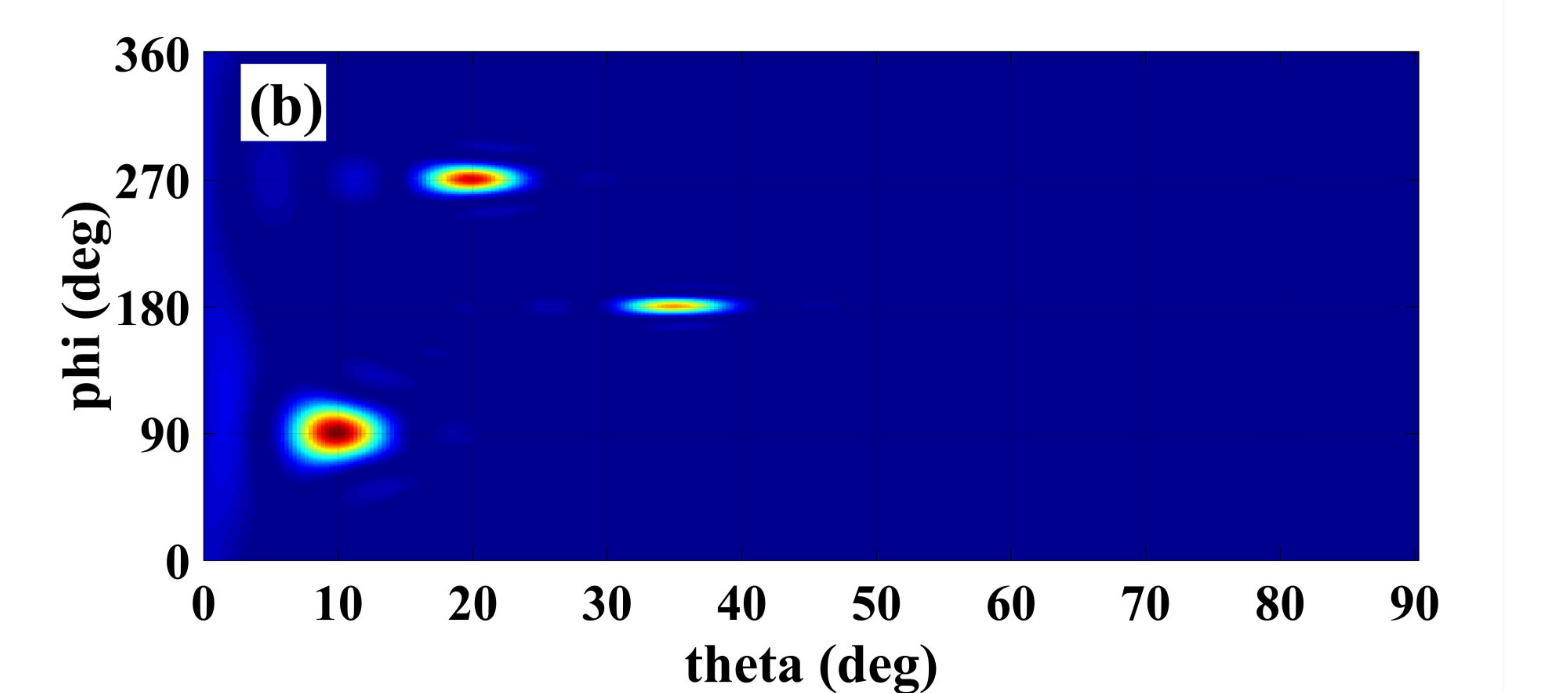} \\
	
	\includegraphics[width=0.5\textwidth]{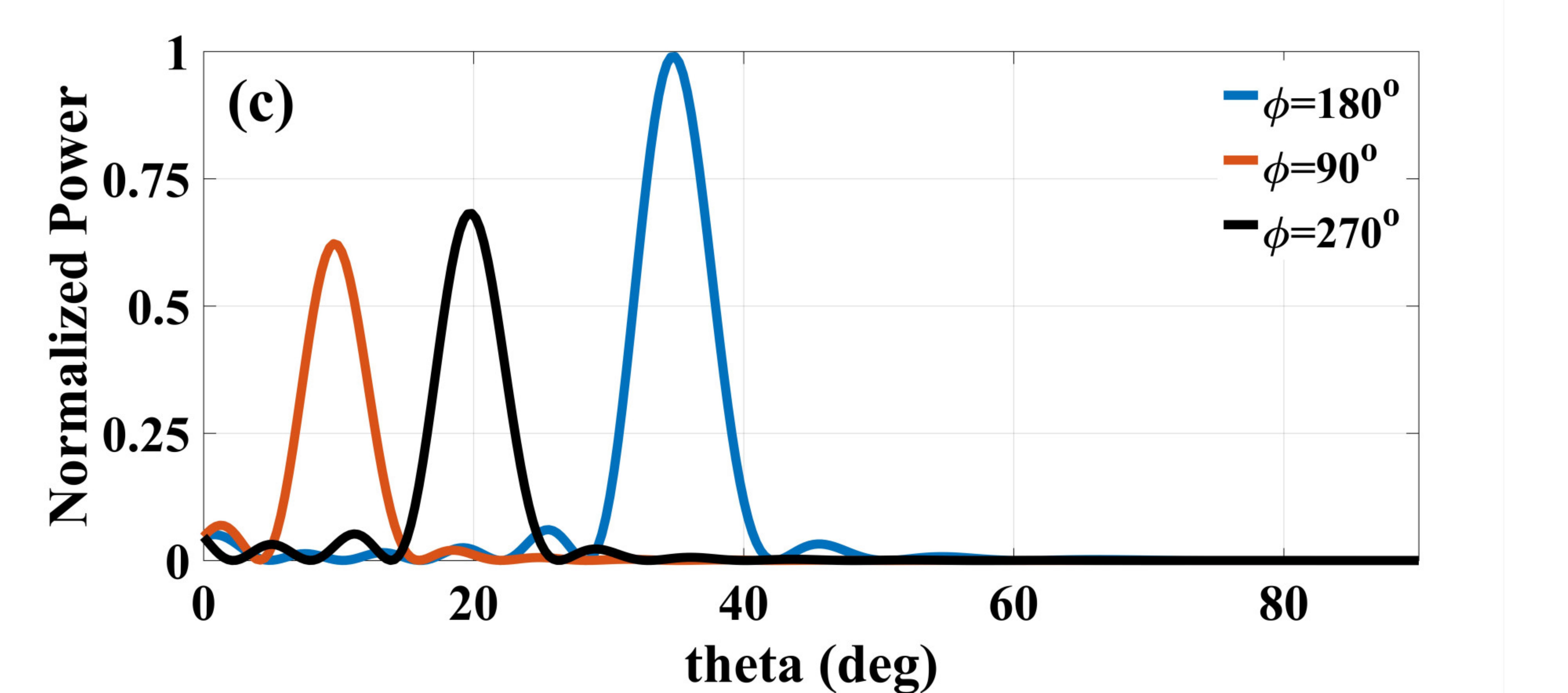}
	\includegraphics[width=0.5\textwidth]{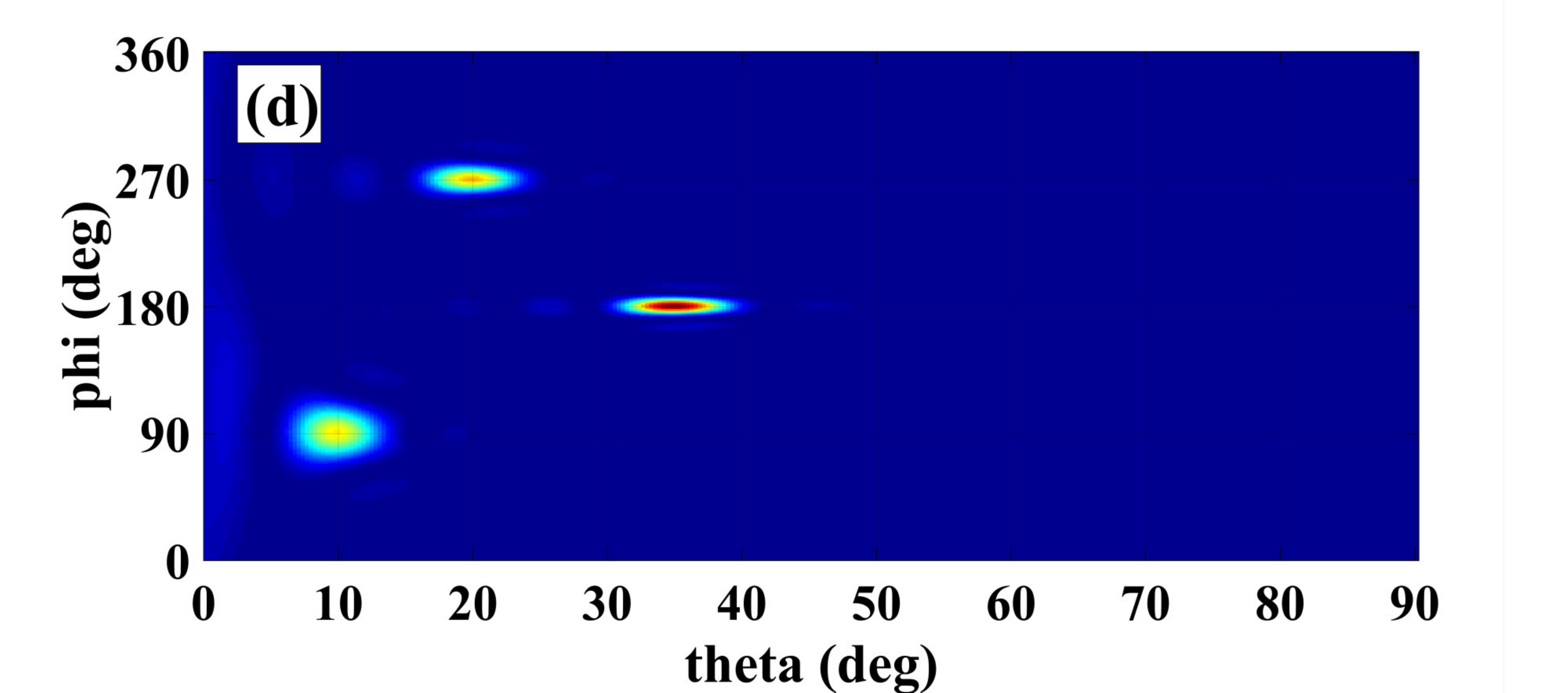} \\
	\includegraphics[width=0.5\textwidth]{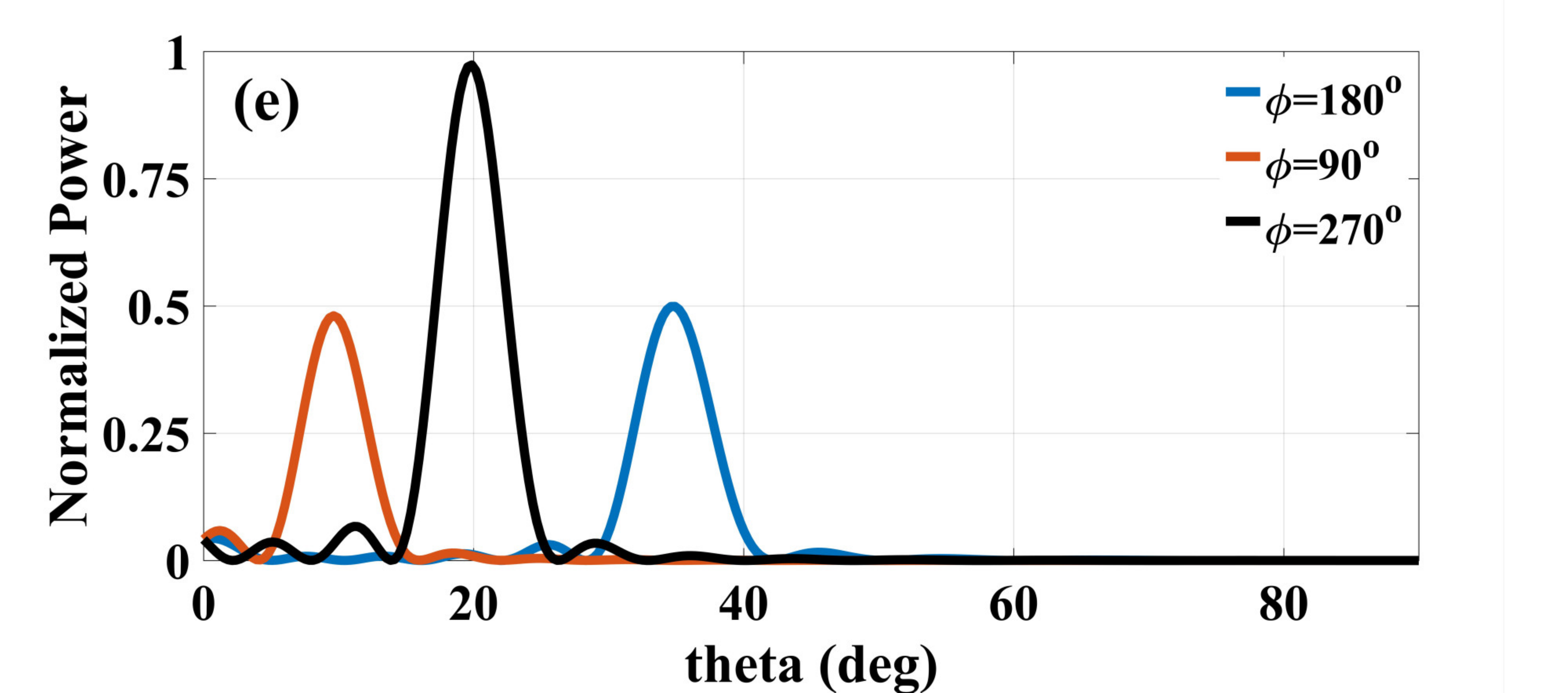}
	\includegraphics[width=0.5\textwidth]{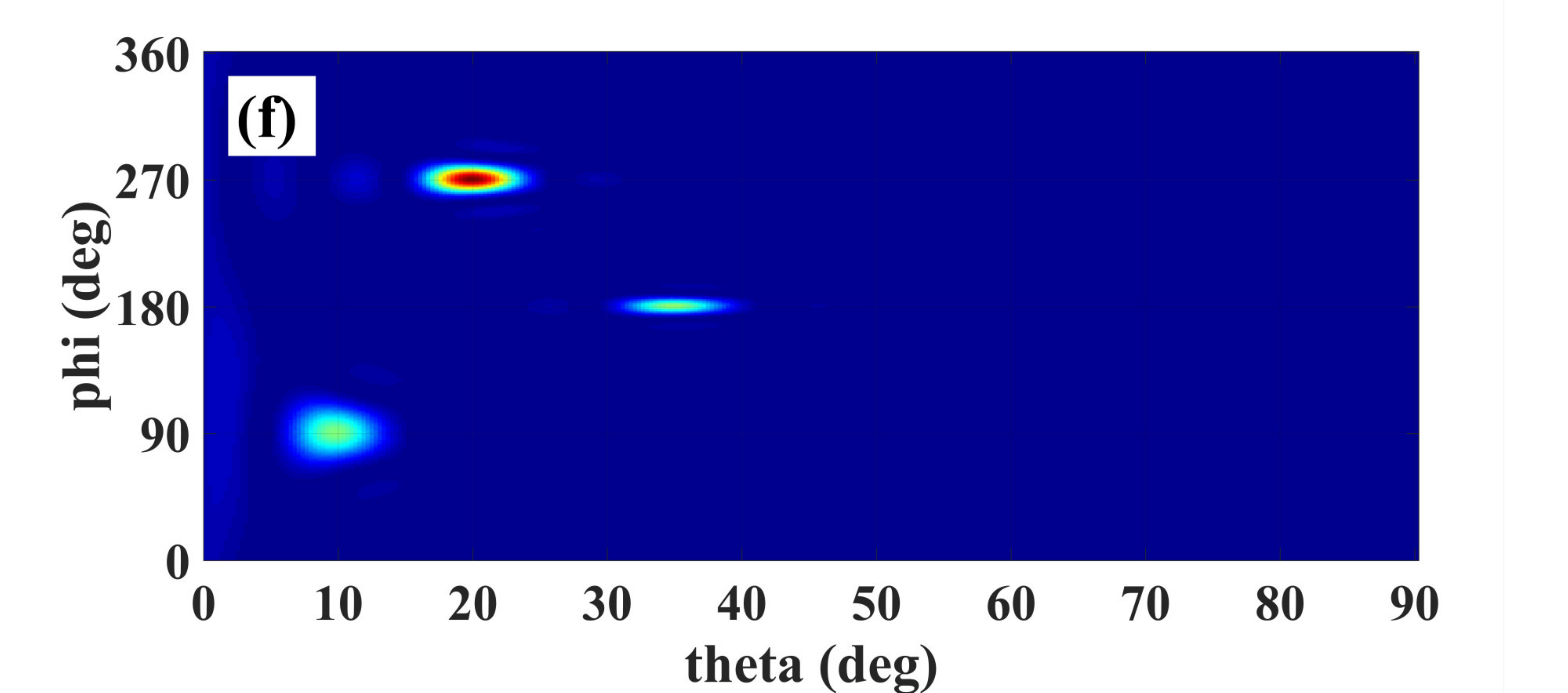}
\end{array}$
\caption[.]{\label{fig:label} Demonstration of three power-dividing examples for different power ratio levels for three beams via monitoring the 2D scattering patterns of the ASPD structure   }
\end{figure*}
 The first coding pattern reflects the incident waves into the direction of $(\theta^{1}=10^\circ,\phi^{1}=180^\circ)$. While, the second one is designed to scatter a single pencil beam pointing at the direction of $(\theta^{2}=30^\circ,\phi^{2}=270^\circ)$. Therefore, the superimposed metasurface (which is driven by $b_{mn}$) obtained from \textcolor{blue}{Eq. 10} and $\sqrt{p^2/p^1}=1$, turns into an ASPD structure that splits the normal incidence into two reflected beams oriented along $\theta^{1}=10^\circ$ and $\theta^{2}=30^\circ$ with the power ratio of $P_{sup}^{gen}(\theta^{2},\phi^{2}){/}P_{sup}^{gen}(\theta^{1},\phi^{1})=0.77$ (see \textcolor{blue}{Figs. 3a,b}). It is worth noting that ignoring the element pattern function of \textcolor{blue}{Eq. 13}, like what is done by \cite{jiu2017flexible,liang2016broadband,wan2016field,liu2017concepts,yan2015broadband} , imports noticeable error in predicting the power ratio levels of differently-oriented pencil beams. For instance, without considering the element pattern function, the superimposed metasurface should generate two pencil beams carrying equal power intensity which essentially contradict with the results of \textcolor{blue}{Figs. 3a,b}. In order to investigate the flexibility of the design, we intend to demonstrate an ASPD dividing the scattered power equally ($P_{sup}^{gen}(\theta^{2},\phi^{2}){/}P_{sup}^{gen}(\theta^{1},\phi^{1})=1$) into two main directions of ($\theta^1  = {10^\circ}\,$ and $\,\theta^2  = {30^\circ}$), a special functionality which the conventional wave-splitting platforms fail to achieve \cite{zhang2018metasurface} . Referring to \textcolor{blue}{Eq. 10}, in order to have two differently-oriented beams with equal power intensities, the ASPD structure must be driven by the superimposed coding pattern governed by the generalized addition theorem where the power coefficients are adopted as $\sqrt{p^2/p^1}=1.137$. As can be seen from \textcolor{blue}{Figs. 3c,d}, thanks to the addition theorem revisiting, the power ratio level of two scattered beams satisfactorily approaches to unity (about 0.99) with the same desired tilt angles. We continue to study another peculiar performance which cannot be realized by traditional coding metasurfaces, i.e. producing two independent asymmetric beams on different planes with the power ratio level of ($P_{sup}^{gen}(\theta^{2},\phi^{2}){/}P_{sup}^{gen}(\theta^{1},\phi^{1})=1.85$) and the orientation of ($\theta^1 = {10^\circ}\,$ and $\,\theta^2 = {30^\circ}$). To attain the additive coding metasurface, we need to perform the revisited version of the addition operation of two coding patterns by setting $\sqrt{p^2/p^1}=1.55$. As the inset of \textcolor{blue}{Figs. 3e,f} demonstrates, 65\% of the reflected power approximately propagates toward the higher elevation angle and the remaining power is arrested in the lower elevation angle, as expected. We remark that the very little discrepancy between our theoretical predictions and MATLAB simulations can be attributed to the unwanted side lobes (information losses) originating from our initial assumptions: continues and laterally infinite phase/amplitude modulating which are not ideally satisfied during the simulations. Therefore, the correctness and robustness of the revisited addition theorem are confirmed theoretically and numerically. The presented revisiting is further surveyed through illustrating different examples in which the superimposed coding metasurfaces act as an ASPD with three scattered beams. In this case, three equations are postulated to disclose the power ratio level between each couple of beams

\begin{equation}
\frac{{{p}^{2}}}{{{p}^{1}}}=\sqrt{\frac{P_{\sup }^{gen}\left( {{\theta }^{2}},{{\phi }^{2}} \right)}{P_{\sup }^{gen}\left( {{\theta }^{1}},{{\phi }^{1}} \right)}}\times \frac{\cos {{\theta }^{1}}}{\cos {{\theta }^{2}}}
\end{equation}
\begin{equation}
\frac{{{p}^{3}}}{{{p}^{2}}}=\sqrt{\frac{P_{\sup }^{gen}\left( {{\theta }^{3}},{{\phi }^{3}} \right)}{P_{\sup }^{gen}\left( {{\theta }^{2}},{{\phi }^{2}} \right)}}\times \frac{\cos {{\theta }^{2}}}{\cos {{\theta }^{3}}}
\end{equation}
\begin{equation}
\frac{{{p}^{3}}}{{{p}^{1}}}=\sqrt{\frac{P_{\sup }^{gen}\left( {{\theta }^{3}},{{\phi }^{3}} \right)}{P_{\sup }^{gen}\left( {{\theta }^{1}},{{\phi }^{1}} \right)}}\times \frac{\cos {{\theta }^{1}}}{\cos {{\theta }^{3}}}
\end{equation}  
           
To continue the concept verification, we wish the ASPD structure to generate three independent pencil beams with $({\theta^1} = {10^\circ},{\varphi^1} = {90^\circ}),({\theta^2} = {20^\circ},{\varphi^2} = {270^\circ}),({\theta^3} = {35^\circ},{\varphi^3} = {180^\circ})$. \textcolor{blue}{Figs. 4a-c} represent  three different power intensity patterns in each of which, three reflected beams with the pre-determined power ratio levels have been successfully acquired by the superimposed coding patterns resulted by $\sqrt{p^1}=\sqrt{p^2}=\sqrt{p^3}=1$ and $\sqrt{p^1}=1$, $\sqrt{p^2}=1.1$, $\sqrt{p^3}=1.5$, and $\sqrt{p^1}=1$, $\sqrt{p^2}=1.48$, $\sqrt{p^3}= 1.2$, respectively. As can be observed from \textcolor{blue}{Figs. 4a-c}, the tilt angles and the power ratios are excellently corroborate our theoretical predictions, i.e.

\begin{tabular}{ c|c|c|c  }
	\multicolumn{4}{c}{} \\ 

	Power Ratio Level    & example\#1 & example\#2 & example\#3 \\
	\hline
$P_{sup}^{gen}(\theta^{2},\phi^{2}){/}P_{sup}^{gen}(\theta^{1},\phi^{1})$ (Theory)& 0.91  & 1.1& 2\\
$P_{sup}^{gen}(\theta^{2},\phi^{2}){/}P_{sup}^{gen}(\theta^{1},\phi^{1})$ (MATLAB simul.)& 0.905  & 1.095& 2.02\\
\hline
$P_{sup}^{gen}(\theta^{3},\phi^{3}){/}P_{sup}^{gen}(\theta^{1},\phi^{1})$ (Theory)& 0.691  & 1.556& 1\\
$P_{sup}^{gen}(\theta^{3},\phi^{3}){/}P_{sup}^{gen}(\theta^{1},\phi^{1})$ (MATLAB simul.)& 0.705  & 1.58& 1.03\\
\hline
$P_{sup}^{gen}(\theta^{3},\phi^{3}){/}P_{sup}^{gen}(\theta^{2},\phi^{2})$ (Theory)& 0.76  & 1.413& 0.5\\
$P_{sup}^{gen}(\theta^{3},\phi^{3}){/}P_{sup}^{gen}(\theta^{2},\phi^{2})$ (MATLAB simul.)& 0.779  & 1.443& 0.509\\

\end{tabular}
\\

With the above discussions, we have made two observations in the revisited version of the addition principle where a laterally-infinite continuously modulated metasurface is under study: (i) deflecting the incident wave into multiple arbitrarily-selected directions and (ii) dividing the power asymmetrically between the radiation beams. Such illustrative examples divulge that theoretically revisiting the addition operations of complex coding patterns via incorporating the amplitude information of the meta-atoms will boost the manipulating abilities of the coding metasurfaces, outstandingly and furnish a robust and flexible design approach for such power-based complicated manipulations. Theoretically speaking, any controllable power ratio levels for multi-beam emissions can be gained by using the revisited version of the addition principle. Prior to providing the meta-atom detail, we stress that in this paper we characterize the meta-atoms occupying the ASPD with both amplitude and phase information. To demonstrate the necessity of such an assumption, let us consider the same approach given in Ref.\cite{wu2018addition} which neglects the reflection amplitude distribution in the superimposed coding patterns. The results of \textcolor{blue}{Fig. 3c} and \textcolor{blue}{Fig. 4c} are re-evaluated but with neglecting the reflection amplitude data.
\begin{figure*}[t!]
	$\begin{array}{rl}
	\includegraphics[width=0.53\textwidth]{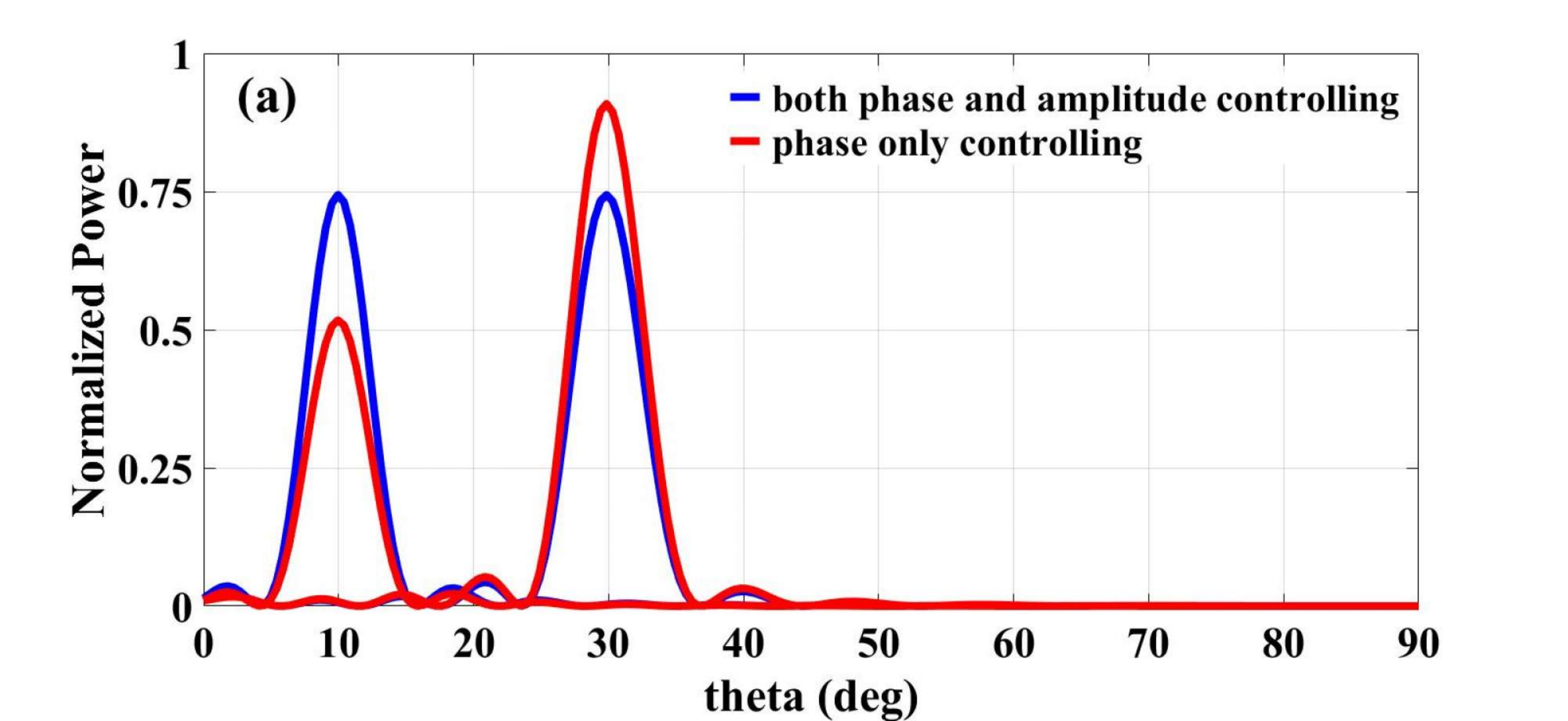} 
	\includegraphics[width=0.53\textwidth]{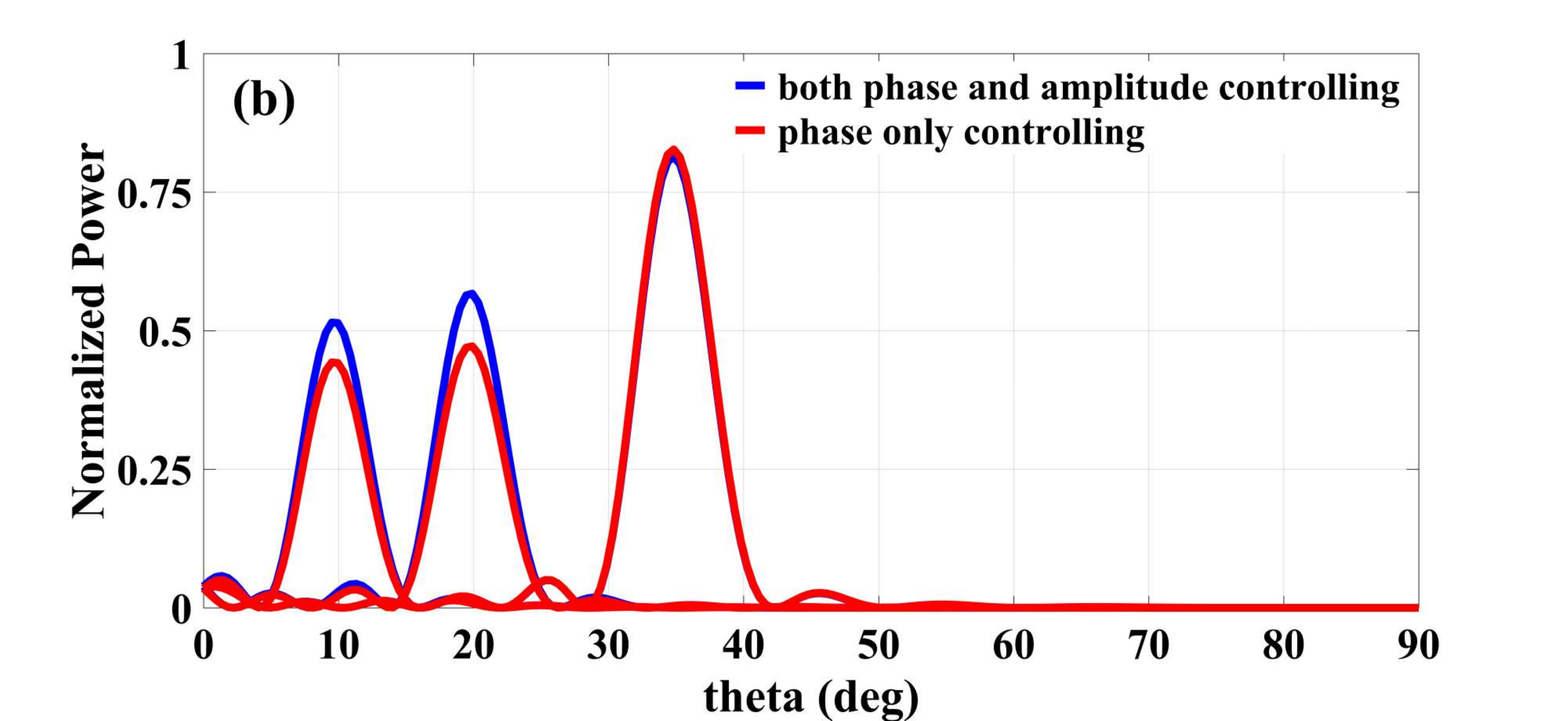} \\
\end{array}$
\caption[.]{\label{fig:label} Highlighting the effects of the amplitude information in the revised version of the addition principle by comparing the 2D scattering patterns of the ASPD designs with/without amplitude data for the results of a) Fig. 3c and b) Fig. 4c. }
\end{figure*}
 As can be seen in \textcolor{blue}{Fig. 5}, the power ratio level of the scattered beams do not further match with our theoretical predictions, thereby, highlighting the significant role of amplitude information in the revisited addition principle. Unlike the previous strategies that are applicable only for predicting the direction of tilted beams, the regulations presented in this study are more general and applicable to study the power intensity patterns of the additive coding metasurfaces. It sounds vital in various antenna applications and simultaneous multi-target detection scenarios that the power ratio levels of differently-oriented scattered beams are required to be manipulated independently.

\section{Numerical Simulations: incorporating discretization and quantization effects}

~~Up to now, we have studied a semi-continues phase/amplitude modulation over a latterly-infinite plane, an optimistic hypothesis which cannot be realized in practice.  A coding metasurface is composed of digital meta-atoms whose size and bit number specify the grade of discretization and quantization, respectively. 
\begin{figure*}[t]
	$\begin{array}{rl}
	\includegraphics[width=0.5\textwidth]{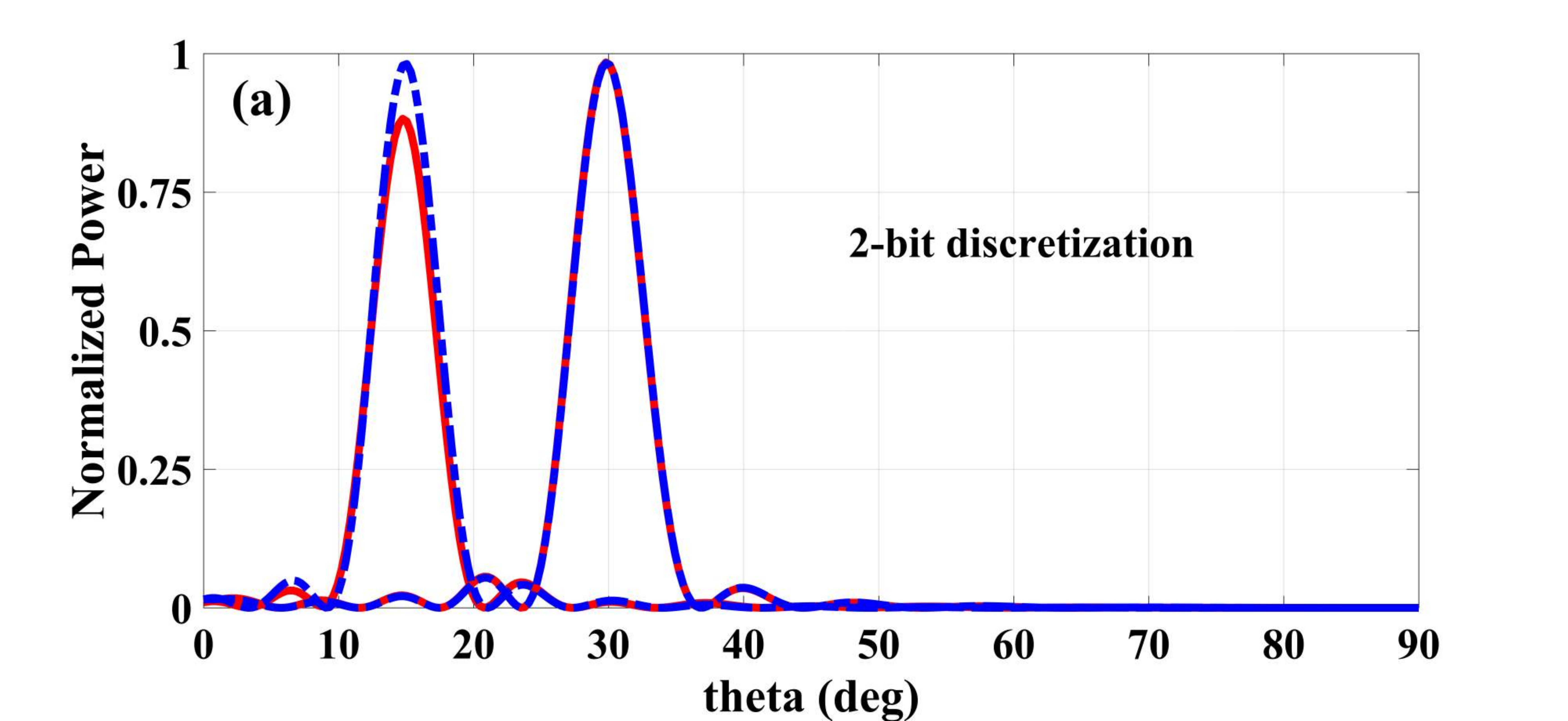} 
	\includegraphics[width=0.5\textwidth]{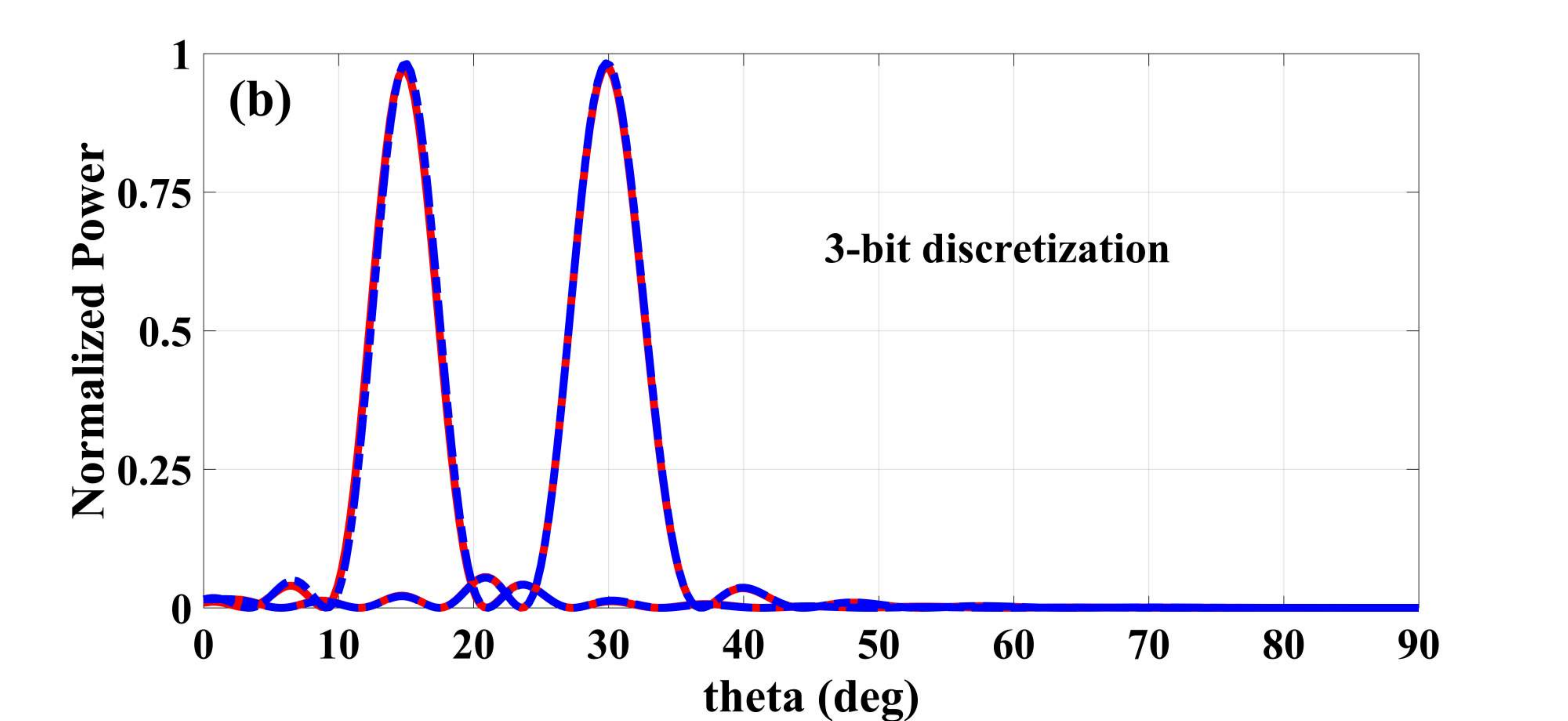} \\
	
	\includegraphics[width=0.5\textwidth]{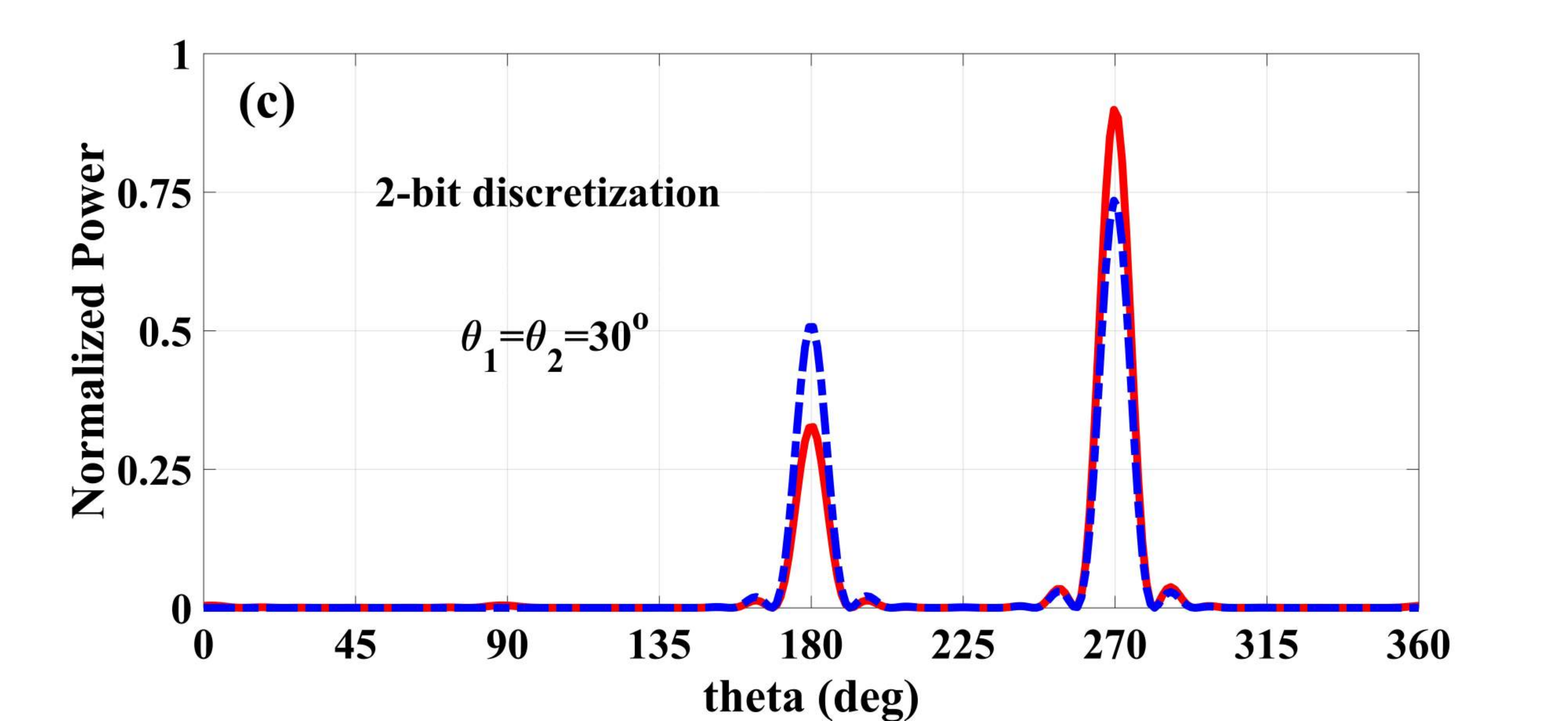}
	\includegraphics[width=0.5\textwidth]{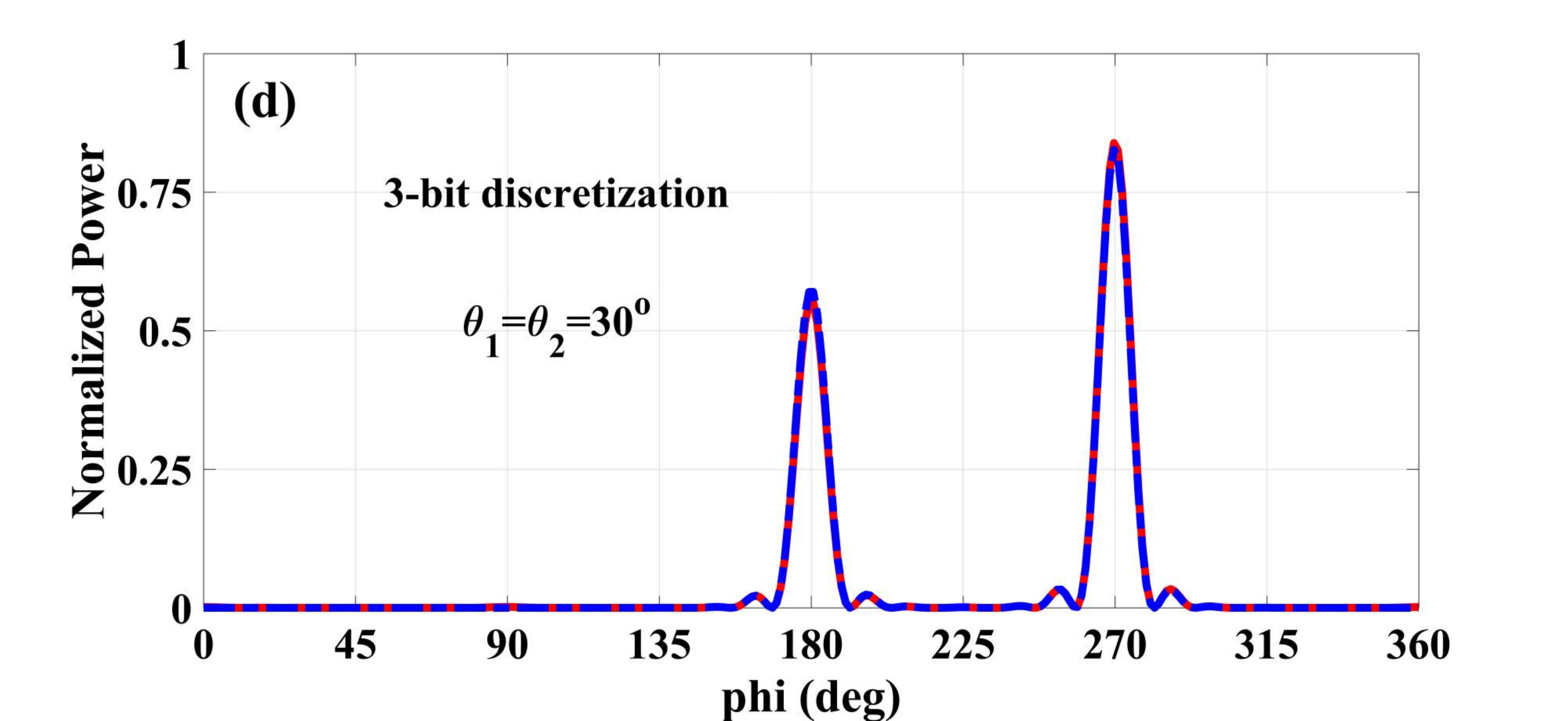} \\
	
\end{array}$
\caption[.]{\label{fig:label} a) and b) are the diagrams for 2-bit and 3-bit discretization process respectively when $({\theta _1} = {15^o},\,{\varphi _1} = {90^o})$ and $({\theta _2} = {30^o},\,{\varphi _2} = {270^o})$. c) and d) are the diagrams for 2-bit and 3-bit discretization process respectively when the two reflected beams are in the same elevation and different azimuth angles. The blue line is the continuous result while the red lines are the discretized results.    }
\end{figure*}
In these wave-manipulating schemes, the discrete reflection phase/amplitude states of the metasurface are adjusted with multi-bit coding sequences realized by spatially inhomogeneous meta-atoms \cite{zhang2017spin,liu2016anomalous,bao2018design} . Since the amplitude/phase profiles of the proposed ASPD structures in \textcolor{blue}{Eq. 6} were assumed ideally continuous in level and spatial position, henceforth, the discretization and quantization effects should be involved in our study, as the constituent meta-atoms have a certain size and phase/amplitude response. In order to investigate the discretization and quantization impacts, the previous simulations are re-accomplished for coding metasurfaces consisting of $N=30$ digitally controlled meta-atoms with the inter-element space of $D_{x}=D_{y}=\lambda/3$ where the phase/amplitude profiles of the superimposed coding pattern are quantized into two and three bits. The 2-bit coding metasurfaces have 16 elements with different phase/amplitude states of “0°/0”, “90°/0.33”, “180°/0.66”, and “270°/1”. Similarly, the building units of the 3-bit coding metasurfaces are characterized with 64 distinct phase/amplitude responses of “0°/0”, “45°/0.14”, “90°/0.28”, “135°/0.42”, “180°/0.57”, “225°/0.71”, “270°/0.85”, and “325°/1”. We continue the study with incorporating the quantization and discretization effects via repeating the antecedent ASPD demonstrations, this time, with 2-bit and 3-bit coding meta-atoms. A fair comparison between the power intensity patterns of the semi-continues and quantized/discretized  amplitude/phase profiles have been carried out and the corresponding results are given in \textcolor{blue}{Figs. 6a-d}. In one of the illustrations, the 2-bit (\textcolor{blue}{Fig. 6a}) and 3-bit (\textcolor{blue}{Fig. 6b}) ASPD structures ($P_{sup}^{gen}(\theta^{2},\phi^{2}){/}P_{sup}^{gen}(\theta^{1},\phi^{1})=1$) are served to divide the incident power between two scattered beams oriented along ($\theta^1=15^\circ,\phi^1=180^\circ$) and ($\theta^2=30^\circ,\phi^2=270^\circ$) directions.  In the other illustration, the 2-bit (\textcolor{blue}{Fig. 6c}) and 3-bit (\textcolor{blue}{Fig. 6d}) ASPD meta-devices ($P_{sup}^{gen}(\theta^{2},\phi^{2}){/}P_{sup}^{gen}(\theta^{1},\phi^{1})=1.44$) are responsible for scattering two pencil beams with the tilt angles of ($\theta^1=30^\circ,\phi^1=180^\circ$) and ($\theta^2=30^\circ,\phi^2=270^\circ$). As can be noticed, although the 2-bit coding architectures fail to achieve satisfactory results in comparison to those of continuously-modulated designs, the 3-bit phase/amplitude-adjustable ASPD structures still correctly and efficiently operate. The quantitative summery of the above-mentioned results is tabulated in \textcolor{blue}{Table. 1}. Consequently, one can deduce that the revisited version of the addition theorem does not remain valid under aggressive quantization levels. Meanwhile, to have a full control over the power pattern intensity of the encoded ASPD designs,  it is required in this paper to modulate both phase and amplitude states of the meta-atoms in 3-bits or higher quantization levels \cite{bao2018design,ding2016dual,farmahini2013metasurfaces,wan2016independent,lee2018complete,zhu2017plasmonic} .

\begin{table}[t]
	\caption{The quantitative comparison between the semi-continues and quantized ASPD designs for spatially dividing the incident power.}
	\label{tbl:example}
	\centering
\begin{tabular}{ c|c|c|c }
	\multicolumn{4}{c}{} \\ 
	Pattern characteristics      & $\frac{P_{sup}^{gen}(\theta^{2},\phi^{2})}{P_{sup}^{gen}(\theta^{1},\phi^{1})}$  & $\frac{P_{sup}^{gen}(\theta^{2},\phi^{2})}{P_{sup}^{gen}(\theta^{1},\phi^{1})}$  &$\frac{P_{sup}^{gen}(\theta^{2},\phi^{2})}{P_{sup}^{gen}(\theta^{1},\phi^{1})}$  \\
	& (2-bit) & (3-bit)&(Theoretical)  \\
	\hline
	$\begin{array}{l}
	({\theta _1} = {15^\circ},\,{\theta _2} = {30^o})\, \,\\
	(\,{\varphi _1} = \,{90^\circ},{\varphi _2} = {0^\circ})
	\end{array}$
	and  $\sqrt{p^2/p^1}=1.115$ & 1.113  & 1& 1\\
	\hline
	$\begin{array}{l}
	({\theta _1} = \,{\theta _2} = {30^\circ})\, \\
	\,({\varphi _1} = \,{180^\circ},\,{\varphi _2}\, = \,{270^\circ})
	\end{array}$
	and  $\sqrt{p^2/p^1}=1.2$ & 2.754  & 1.45& 1.44\\
\end{tabular}
\end{table}
Different methodological attempts have been made to simultaneously modulate the amplitude and phase profiles of a metasurface via tuning the geometry of antennas \cite{bao2018design} . The meta-atoms employed in this paper integrate the functionality of a metasurface for phase control and a metasurface for amplitude control which are adjusted with the geometrical configuration and angular orientation of C-shaped particles, respectively\cite{bao2018design,liu2016anomalous} . The phase-controlling metasurface is composed of geometrically-engineered  C-shaped antennas and functions in a linear cross-polarization scheme. Indeed, the phase can be robustly and independently controlled in the reflection mode with a polarization orthogonal to that of the incident wave by the arm length and the open angle (\textcolor{blue}{Fig. 1b}). The symmetry line of each particle is oriented along +45° or –45° angle to maximize the polarization conversion ratio merit \cite{bao2018design} .
 \begin{figure*}[t]
	$\begin{array}{rl}
	\includegraphics[width=0.5\textwidth]{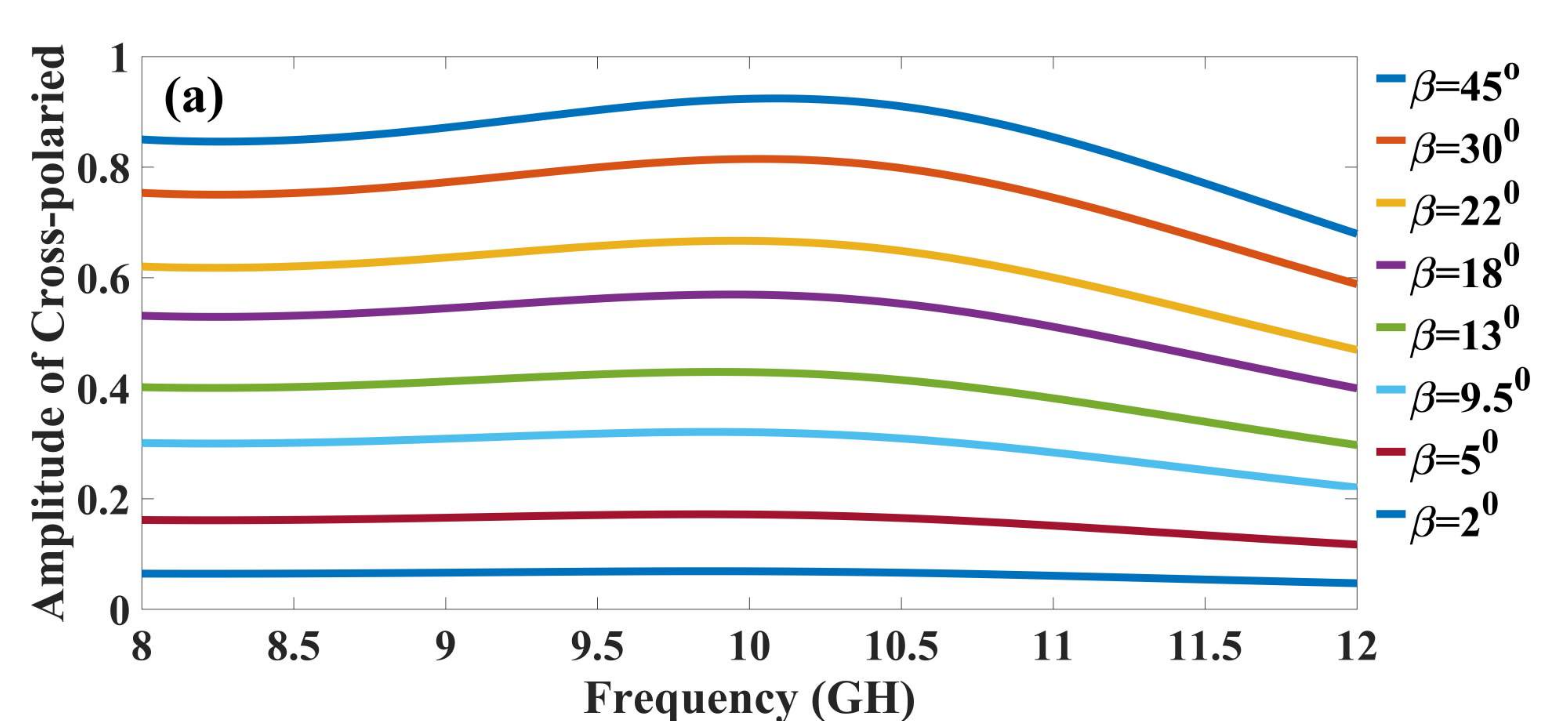} 
	\includegraphics[width=0.5\textwidth]{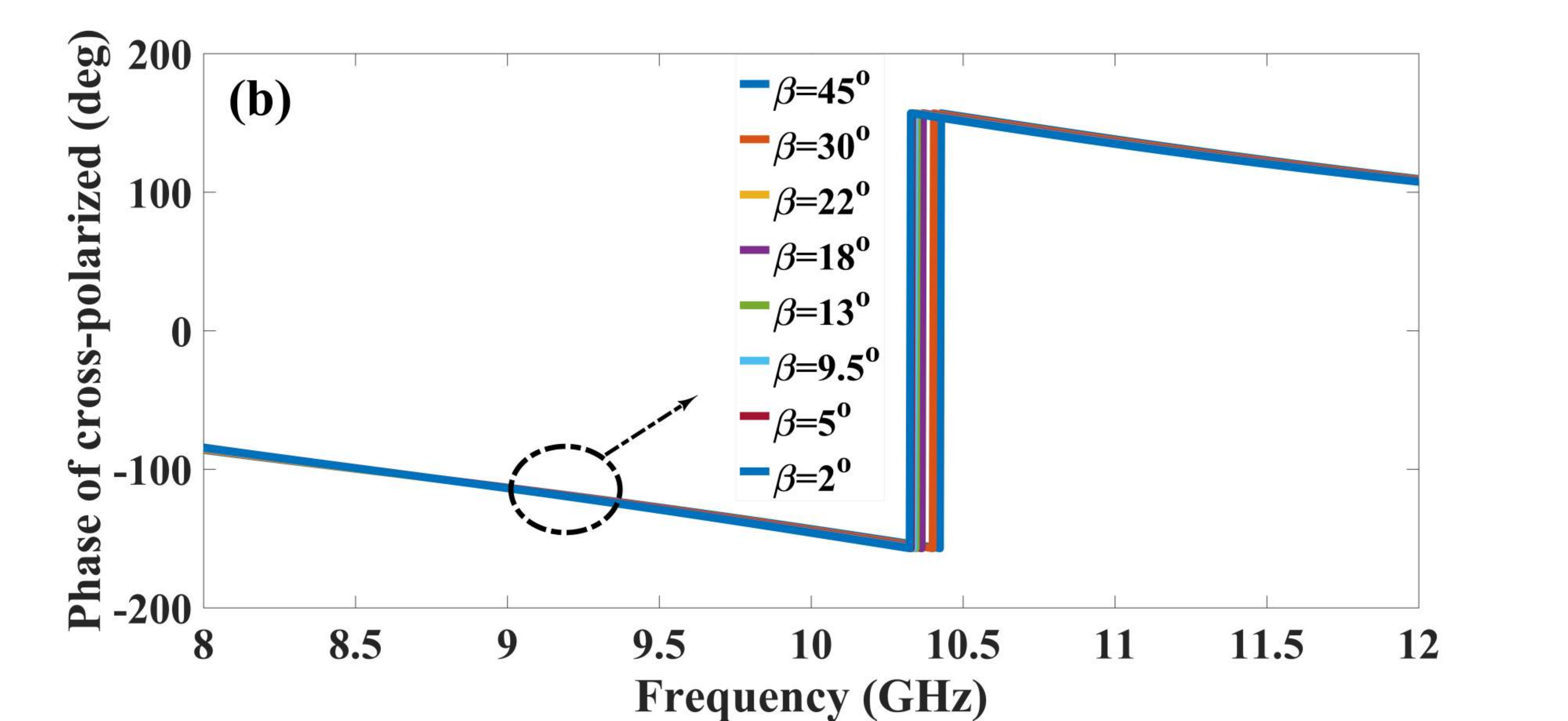} \\
	\includegraphics[width=0.5\textwidth]{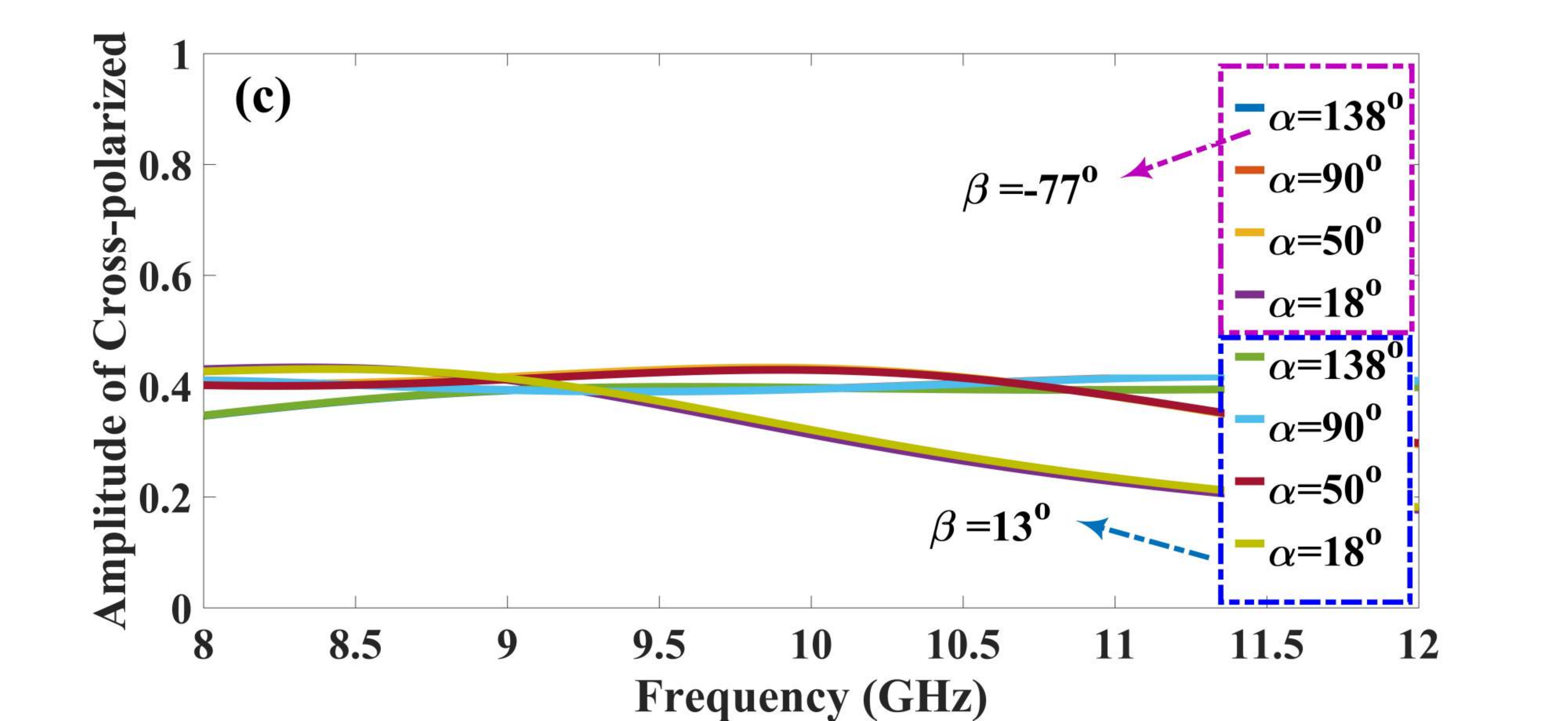}
	\includegraphics[width=0.5\textwidth]{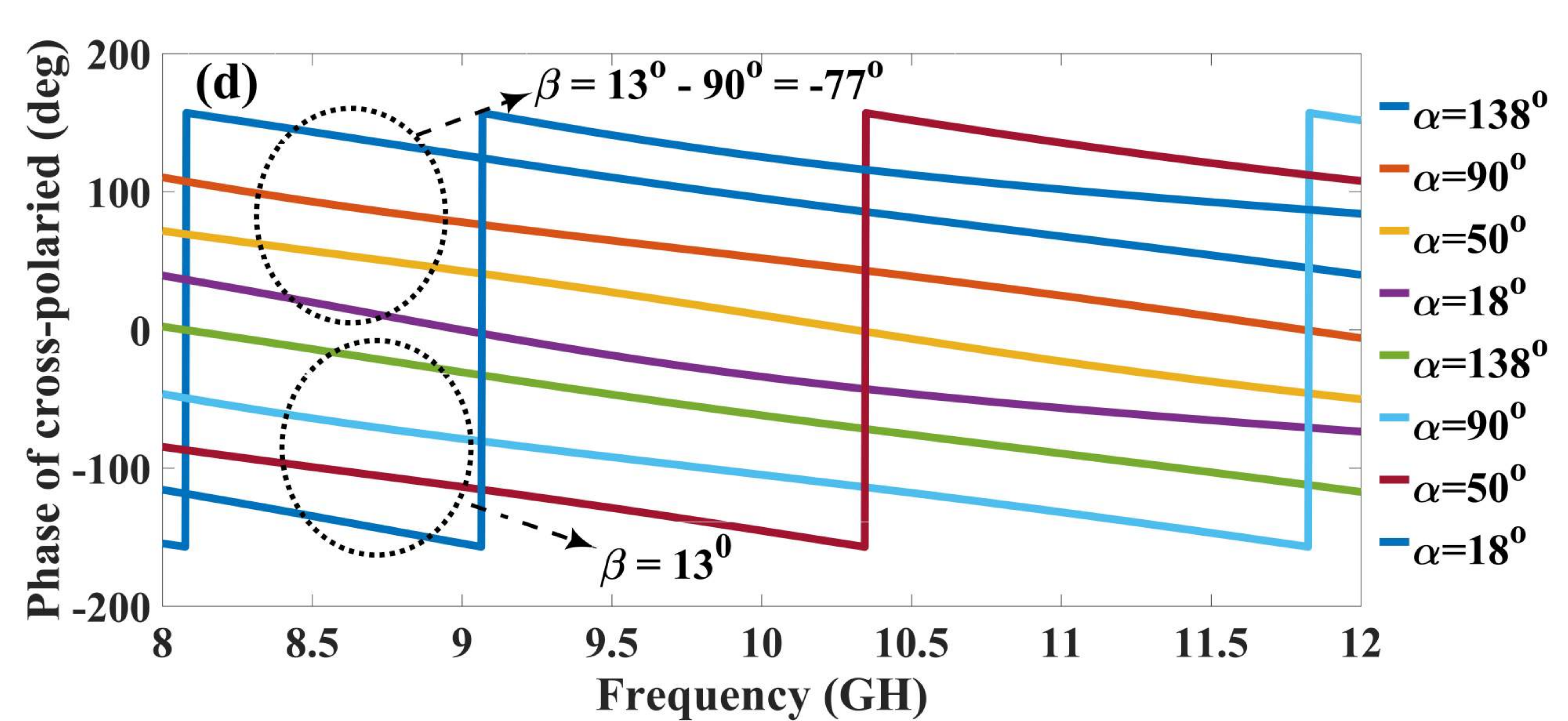} \\ 
	\end{array}$
	\caption[.]{\label{fig:label} The simulated a) amplitude and b) phase of the C-shaped meta-atom with different $\beta $ values when $\alpha  = {50^\circ}$. The simulated c) amplitude and d) phase of the C-shaped meta-atom with different $\alpha$ values when $\beta  = {13^circ}$ and $\beta  = {-77^\circ}$   }
\end{figure*}
 The amplitude-controlling metasurface, however, consists of C-shaped meta-atoms of the same geometry in which the reflection amplitude can be independently modulated with varying the orientations\cite{bao2018design,zhang2017spin,liu2016anomalous} . Merging the design rules of these two types of metasurfaces yields a flexible C-shaped meta-atom whose phase and amplitude responses can be separately manipulated by changing its geometry and orientation (\textcolor{blue}{Figs. 7a-d}). The front view of the established meta-atoms responsible for locally and separately tuning the amplitude and phase of the reflected cross-polarized wave is pictured in \textcolor{blue}{Fig. 1b}. The meta-atom is composed of C-shaped metallic structure etched on a 3.2mm-thick FR4 substrate ($\epsilon_r$=4.3 and $\tan\delta$=0.025). To block the transmitted power, the structure is terminated with a copper ($\sigma  = 5.8 \times {10^7}\,$S/m) ground plane. The periodicity of meta-atom is equal to $\lambda /3$ where $\lambda$ is the operating wavelength at $10$GHz and the other geometrical parameters are given in the caption of \textcolor{blue}{Fig. 1}. Through engineering the geometrical parameters of open angle ($\alpha$) and orientation angle ($\beta$), one can independently control the phase and amplitude of the cross-polarized reflection, respectively, a vital constraint to realize the revisited version of the addition theorem. The proposed meta-atoms are characterized with the Floquet solver of the commercial software CST Microwave Studio where periodic boundary conditions are applied to x- and y-directions to simulate a latterly-infinite periodic array while the Floquet ports are assigned to the z-directin. \textcolor{blue}{Figs. 7a,b} and \textcolor{blue}{Figs. 7c,d} demonstrate the simulated cross-polarized reflection phase and amplitude coefficients with varying $\beta$ and $\alpha$ parameters, respectively.
  \begin{figure*}[t]
 	$\begin{array}{rl}
 	\includegraphics[width=0.8\textwidth]{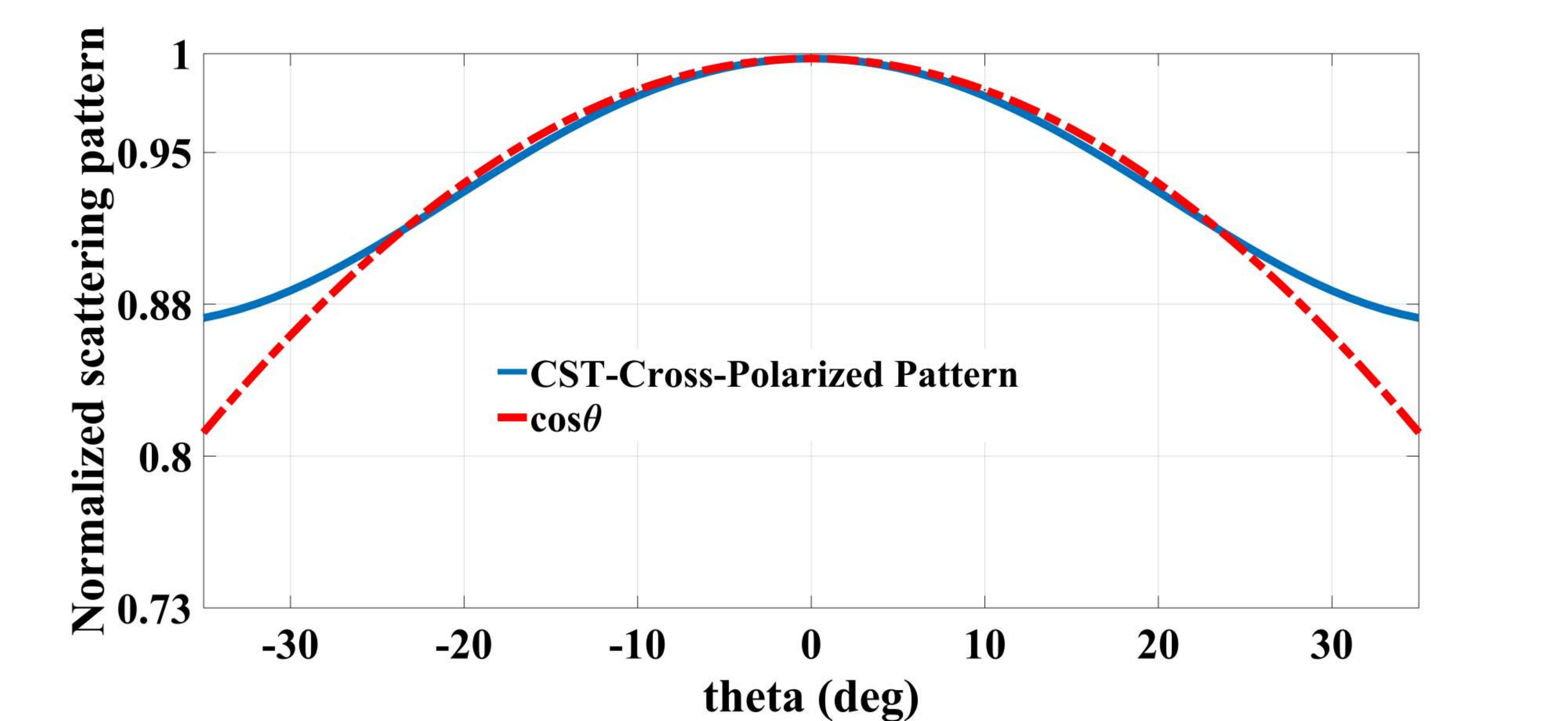} 
 	\end{array}$
 	\caption[.]{The simulated scattering pattern function of the employed meta-atoms.  }
 \end{figure*}
 When $\beta \, = \, \pm \,{45^\circ}$, the maximum energy from the incident wave will be coupled to the cross-polarization component and for the case of $\beta$=${0^\circ}$ or  ${90^\circ}$, the cross-polarization component will vanish. Meanwhile, the cross-polarized reflection phase spans whole ${180^\circ}$ phase range when $\alpha$ varies and $\beta$ is kept constant. By adding $ \pm \,{90^o}$  to $\beta $, the amplitude remains constant while the reflection phase experiences further changes of $ \pm \,{180^\circ}$ . Eventually, a robust and facile approach is investigated for achieving simultaneous phase and amplitude manipulation in a single layer metasurface over a broadband frequency range in the microwave regime \cite{liu2014broadband} . The pattern function of the meta-atom employed in this paper (see \textcolor{blue}{Fig. 1b}) is also simulated in CST Microwave Studio and the results are displayed in \textcolor{blue}{Fig. 8}. This figure illustrates that the designed meta-atoms have maximum scattering field intensity along the boresight direction while it falls down gradually when observation direction deviates from $\theta=0$.  Hereafter, this element factor which can be approximated by cosine function will be involved in \textcolor{blue}{Eq. 12} to accurately predict the power ratio level of multiple beams scattered by the ASPD designs. 

  \begin{figure}[t]
	$\begin{array}{rl}
	\includegraphics[width=0.32\textwidth]{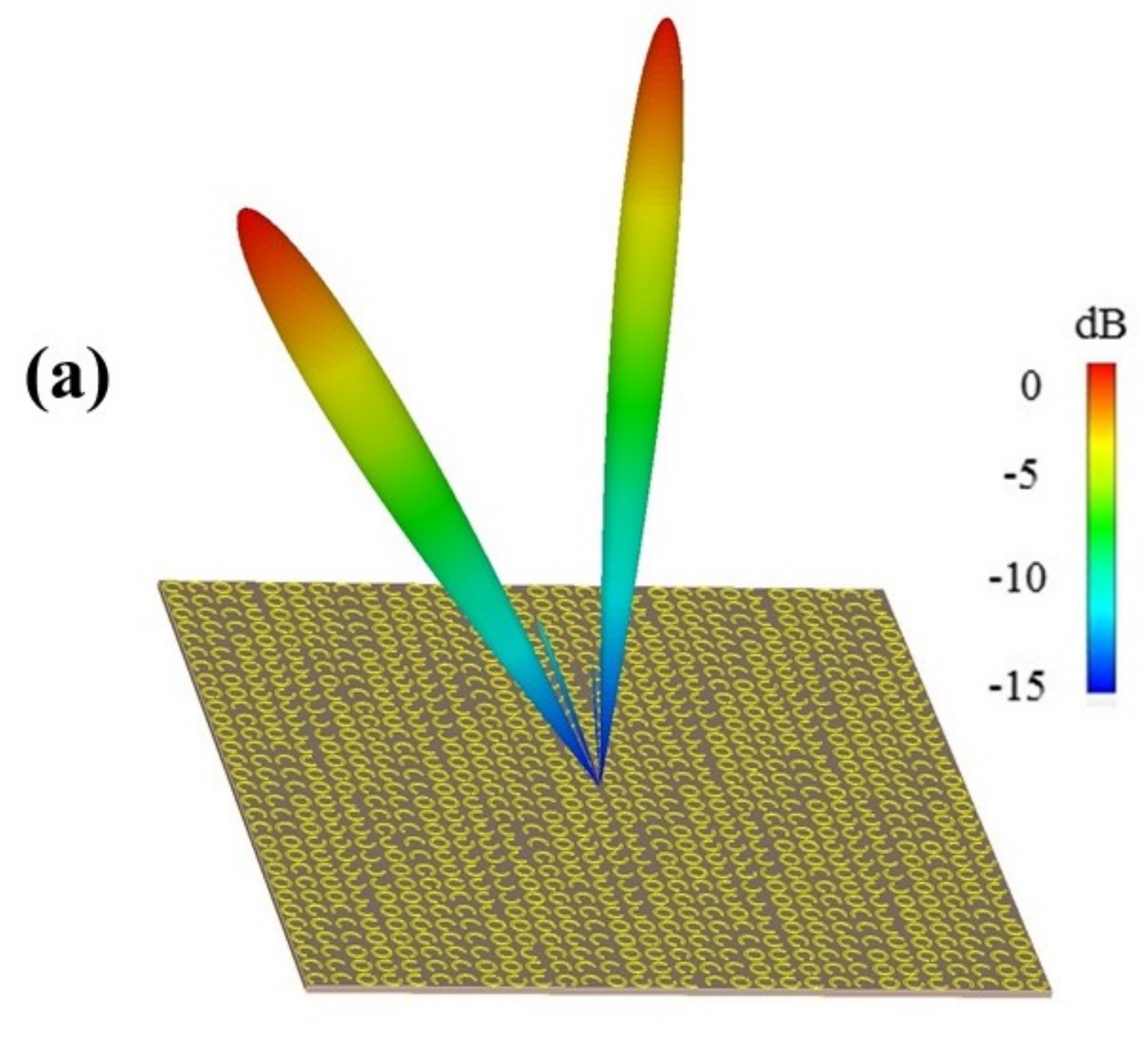} 
	\includegraphics[width=0.5\textwidth]{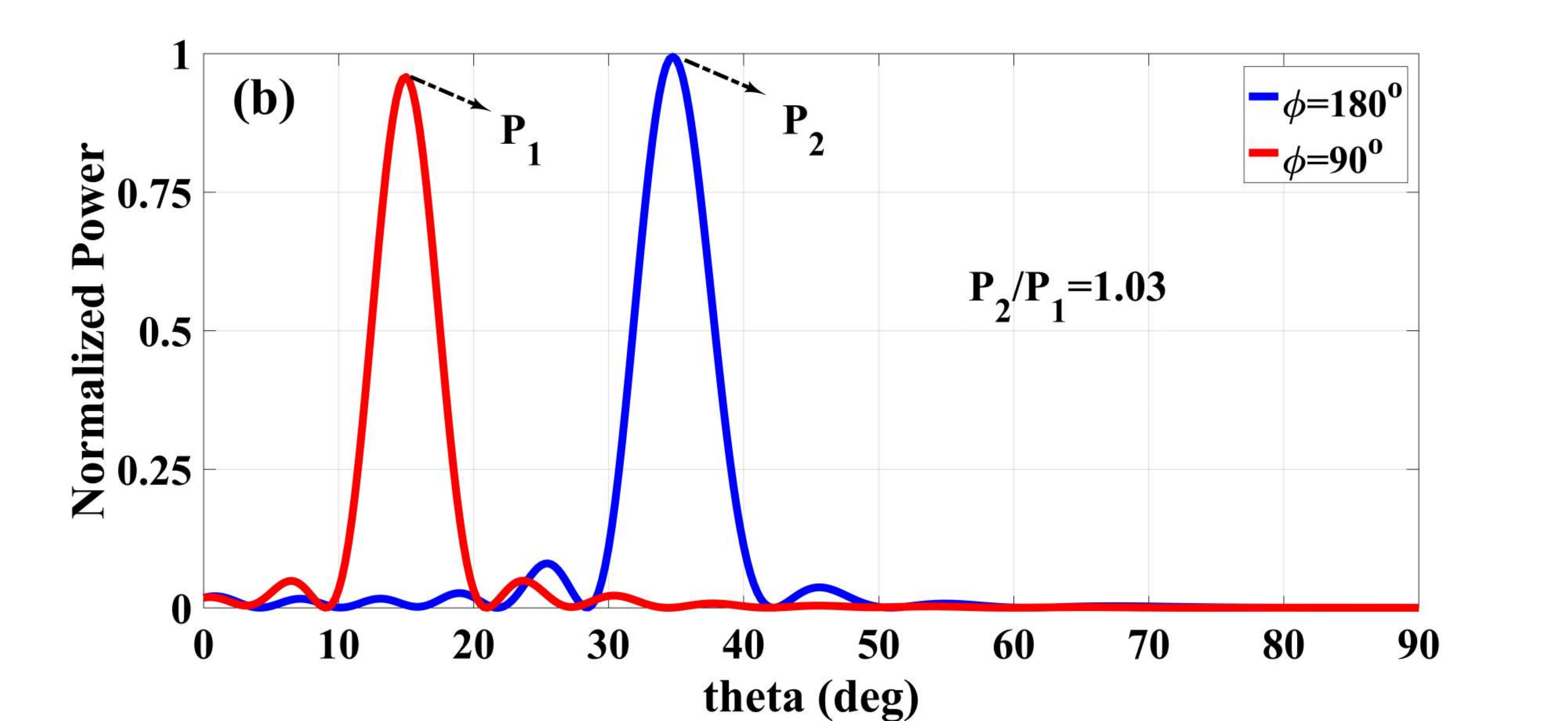} \\
	\includegraphics[width=0.32\textwidth]{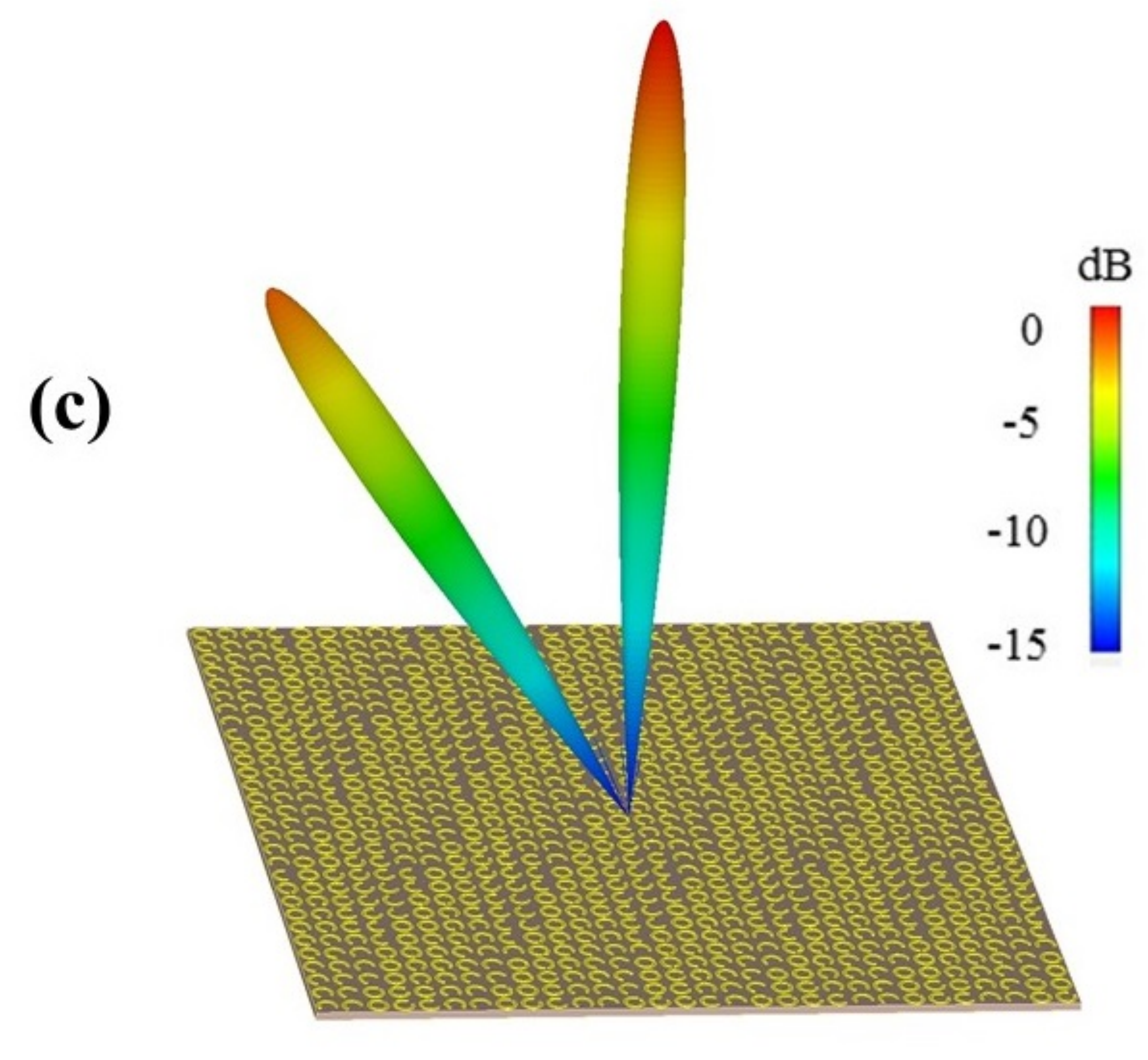} 
	\includegraphics[width=0.5\textwidth]{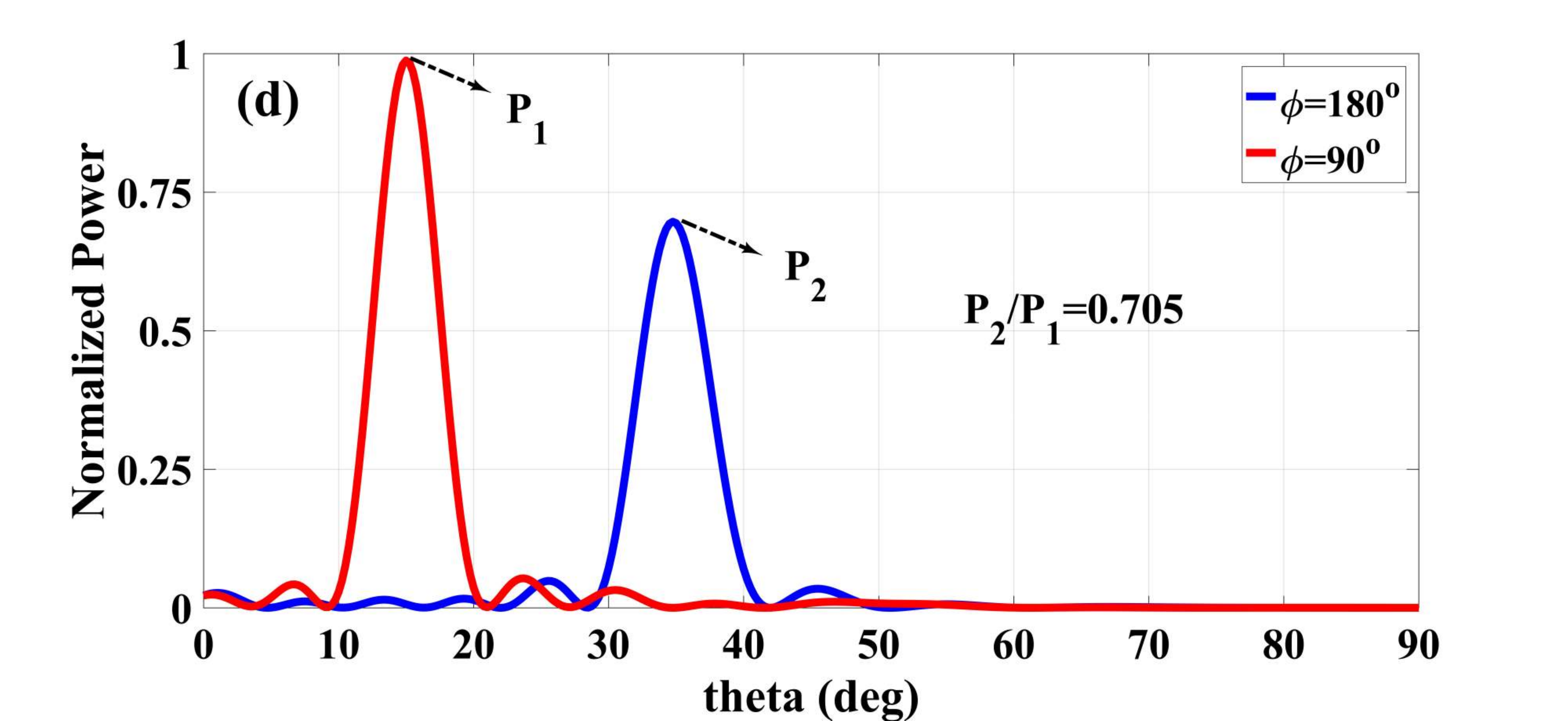} \\
	\end{array}$
	\caption[.]{\label{fig:label} The simulated a) 3D and b) 2D scattering patterns of the ASPD structures responsible for a,b) equally and c,d) unequally dividing the incident power between two beams.}
\end{figure}
\begin{table}[t]
	\caption{The quantitative comparison between the full-wave simulations and theoretical predictions for the power ratio levels of the encoded ASPD designs.}
	\label{tbl:example}
	\centering
	\begin{tabular}{ c|c|c }
		\multicolumn{3}{c}{} \\ 
		Pattern characteristics      & $\frac{P_{sup}^{gen}(\theta^{2},\phi^{2})}{P_{sup}^{gen}(\theta^{1},\phi^{1})}$  & $\frac{P_{sup}^{gen}(\theta^{2},\phi^{2})}{P_{sup}^{gen}(\theta^{1},\phi^{1})}$    \\
		& (Theoretical)  & (Numerical) \\
		\hline
		$\begin{array}{l}
		({\theta _1} = {15^\circ},{\theta _2} = {35^\circ})\,\\
		\,({\varphi _1} ={90^\circ}, {\varphi _2} = {180^\circ})
		\end{array}$
		and  $\sqrt{p^2/p^1}=1.179$ & 1  &  1.03\\
		\hline
		$\begin{array}{l}
		({\theta _1} = {15^\circ},{\theta _2} = {35^\circ})\,\\
		\,({\varphi _1} = {90^\circ},{\varphi _2} = {180^\circ})
		\end{array}$ 
		and  $\sqrt{p^2/p^1}=1$ & 0.719  & 0.705\\
	\end{tabular}
\end{table}
To further validate the concept and dive in the performance of our designs in a more realistic configurations, the finite-size ASPD architectures are excited by a normal plane wave in a full-wave simulation host, CST Microwave Studio. The design follows our previously developed analytical formalism and we characterize the functionality of the phase/amplitude-encoded ASPD (realized by 3-bit coding meta-atoms) through a bistatic scattering pattern measurement setup, numerically. In the following, two specific examples are perused by simulations to show the powerful ability of the proposed ASPD meta-devices in generating multi-beam scattered beams with asymmetric power ratio levels. Without loss of generality, the y-polarized normal incidence is considered here. As the first demonstration, aided by the addition theorem revisiting, we have combined two gradient metasurface producing pencil beams along $({\theta _1} = \,{15^\circ},{\varphi _1}\, ={90^\circ})$ and $(\,{\theta _2}\, = {35^\circ},{\varphi _2}\, = {180^\circ})$ directions, respectively, with identical power coefficients of $(\sqrt{p^2/p^1}= 1.179)$. Therefore, the superimposed metasurface should emit two main beams whose carried powers are dictated by their tilt angles (see \textcolor{blue}{Figs. 9a,b}). As expected from the revised addition principle, \textcolor{blue}{Eq. 12}, the power ratio of two scattered beams must obey $P_{sup}^{gen}(\theta^{2},\phi^{2}){/}P_{sup}^{gen}(\theta^{1},\phi^{1})=1$. The full wave simulation results of the 3D and 2D scattering patterns depicted in \textcolor{blue}{Figs. 9a,b} corroborate well our theoretical prediction where they report the power ratio level of beams as 1.03 and our desired beam angles as $\theta^1=15^\circ$ and $\theta^2=35^\circ$. The slight discrepancy between the results can be attributed to the discretization effects, finite size of the metasurface and the inevitable coupling between the meta-atoms. It should be noted that in the fist illustration, the difference between the power intensity levels are just commanded by thier distinct tilt angles. For the second demonstration, the basic gradient metasurfaces creating two pencil beams pointing at $({\theta _1} = \,{15^\circ},{\varphi _1}\, ={90^\circ})$  and 
$(\,{\theta _2}\, = {35^\circ},{\varphi _2}\, = {180^\circ})$ directions are added by choosing the power coefficients of $\sqrt{p^2/p^1}= 1$. By applying the power coefficients into the revised addition theorem, the scattered beams are noticed to have the power ratio level of $P_{sup}^{gen}(\theta^{2},\phi^{2}){/}P_{sup}^{gen}(\theta^{1},\phi^{1})=0.719$. The simulated results of the 2D and 3D bistatic scattering patterns (see \textcolor{blue}{Figs. 9c,d}) show that the power ratio is 0.705 that has an excellent conformity with our analytical predictions. Moreover, the scattered beams are satisfactorily oriented along the pre-determined directions. The summery of the above results are listed in \textcolor{blue}{Table. 2}. The quantitative achievements resulted by the comparison between analytical and numerical results, depict a perfect concordance. The existing negligible errors are mostly due to the discretization and finite size of the ASPD which are interestingly less than 3\%. In overall, the proposed ASPD structures successfully perform their missions, i.e. dividing the power asymmetrically and arbitrarily between multiple beams pointing at our desired directions, an outstanding functionality which was not reported for the coding metasurfaces.       

\section{Conclusion}
~~~To sum up, according to this point that the ability in controlling the power distribution of scattered beams is highly demanded by engineers in diverse practical application, the emptiness of exploiting a straightforward and fast way to meet the existing requirements for coding metasurfaces was severely sensed. To fill this vacancy, by revisiting the superposition principle, a revised version of the addition theorem was proposed for the first time to estimate the exact amount of the power ratio of the multiple beams. As a consequence, we introduced the concept of ASPD architectures whereby the power can be spatially, asymmetrically and arbitrarily divided between multiple beams oriented along the pre-determined directions. The proposed design scheme is not accompanied with any brute-force optimization, trial-and-error steps, and time-consuming procedures. Particularly, it was theoretically shown that unlike the previous demonstrations, both phase and amplitude profiles of the coding metasurfaces must be modulated to empower us to flexibly control their power intensity patterns. Benefited from C-shaped meta-atoms, we was able to independently tailor the local reflection phase and amplitude in a 3-bit quantization level. Several illustrative examples were presented in this paper to validate the concept. This work takes a great step forward in designing spatial power dividers for which many promising applications such as beamforming networks and MIMO communication can be envisioned.





\bibliography{achemso-demo}

\providecommand{\latin}[1]{#1}
\makeatletter
\providecommand{\doi}
  {\begingroup\let\do\@makeother\dospecials
  \catcode`\{=1 \catcode`\}=2 \doi@aux}
\providecommand{\doi@aux}[1]{\endgroup\texttt{#1}}
\makeatother
\providecommand*\mcitethebibliography{\thebibliography}
\csname @ifundefined\endcsname{endmcitethebibliography}
  {\let\endmcitethebibliography\endthebibliography}{}
\begin{mcitethebibliography}{58}
\providecommand*\natexlab[1]{#1}
\providecommand*\mciteSetBstSublistMode[1]{}
\providecommand*\mciteSetBstMaxWidthForm[2]{}
\providecommand*\mciteBstWouldAddEndPuncttrue
  {\def\EndOfBibitem{\unskip.}}
\providecommand*\mciteBstWouldAddEndPunctfalse
  {\let\EndOfBibitem\relax}
\providecommand*\mciteSetBstMidEndSepPunct[3]{}
\providecommand*\mciteSetBstSublistLabelBeginEnd[3]{}
\providecommand*\EndOfBibitem{}
\mciteSetBstSublistMode{f}
\mciteSetBstMaxWidthForm{subitem}{(\alph{mcitesubitemcount})}
\mciteSetBstSublistLabelBeginEnd
  {\mcitemaxwidthsubitemform\space}
  {\relax}
  {\relax}

\bibitem[Engheta and Ziolkowski(2006)Engheta, and
  Ziolkowski]{engheta2006metamaterials}
Engheta,~N.; Ziolkowski,~R.~W. \emph{Metamaterials: physics and engineering
  explorations}; John Wiley \& Sons, 2006\relax
\mciteBstWouldAddEndPuncttrue
\mciteSetBstMidEndSepPunct{\mcitedefaultmidpunct}
{\mcitedefaultendpunct}{\mcitedefaultseppunct}\relax
\EndOfBibitem
\bibitem[Holloway \latin{et~al.}(2012)Holloway, Kuester, Gordon, O'Hara, Booth,
  and Smith]{holloway2012overview}
Holloway,~C.~L.; Kuester,~E.~F.; Gordon,~J.~A.; O'Hara,~J.; Booth,~J.;
  Smith,~D.~R. An overview of the theory and applications of metasurfaces: The
  two-dimensional equivalents of metamaterials. \emph{IEEE Antennas and
  Propagation Magazine} \textbf{2012}, \emph{54}, 10--35\relax
\mciteBstWouldAddEndPuncttrue
\mciteSetBstMidEndSepPunct{\mcitedefaultmidpunct}
{\mcitedefaultendpunct}{\mcitedefaultseppunct}\relax
\EndOfBibitem
\bibitem[Chu \latin{et~al.}(2018)Chu, Li, Liu, Luo, Sun, Hang, Zhou, and
  Lai]{chu2018hybrid}
Chu,~H.; Li,~Q.; Liu,~B.; Luo,~J.; Sun,~S.; Hang,~Z.~H.; Zhou,~L.; Lai,~Y. A
  hybrid invisibility cloak based on integration of transparent metasurfaces
  and zero-index materials. \emph{Light: Science \& Applications}
  \textbf{2018}, \emph{7}, 50\relax
\mciteBstWouldAddEndPuncttrue
\mciteSetBstMidEndSepPunct{\mcitedefaultmidpunct}
{\mcitedefaultendpunct}{\mcitedefaultseppunct}\relax
\EndOfBibitem
\bibitem[Zhu \latin{et~al.}(2013)Zhu, Feng, Zhang, Yin, and Zhang]{zhu2013one}
Zhu,~X.; Feng,~L.; Zhang,~P.; Yin,~X.; Zhang,~X. One-way invisible cloak using
  parity-time symmetric transformation optics. \emph{Optics letters}
  \textbf{2013}, \emph{38}, 2821--2824\relax
\mciteBstWouldAddEndPuncttrue
\mciteSetBstMidEndSepPunct{\mcitedefaultmidpunct}
{\mcitedefaultendpunct}{\mcitedefaultseppunct}\relax
\EndOfBibitem
\bibitem[Kim \latin{et~al.}(2018)Kim, An, Cho, Hyun, Moon, Kim, and
  Park]{kim2018full}
Kim,~S.-W.; An,~B.~W.; Cho,~E.; Hyun,~B.~G.; Moon,~Y.-J.; Kim,~S.-K.;
  Park,~J.-U. A Full-Visible-Spectrum Invisibility Cloak for Mesoscopic Metal
  Wires. \emph{Nano letters} \textbf{2018}, \relax
\mciteBstWouldAddEndPunctfalse
\mciteSetBstMidEndSepPunct{\mcitedefaultmidpunct}
{}{\mcitedefaultseppunct}\relax
\EndOfBibitem
\bibitem[Smith \latin{et~al.}(2004)Smith, Pendry, and
  Wiltshire]{smith2004metamaterials}
Smith,~D.~R.; Pendry,~J.~B.; Wiltshire,~M.~C. Metamaterials and negative
  refractive index. \emph{Science} \textbf{2004}, \emph{305}, 788--792\relax
\mciteBstWouldAddEndPuncttrue
\mciteSetBstMidEndSepPunct{\mcitedefaultmidpunct}
{\mcitedefaultendpunct}{\mcitedefaultseppunct}\relax
\EndOfBibitem
\bibitem[Zhang \latin{et~al.}(2009)Zhang, Park, Li, Lu, Zhang, and
  Zhang]{zhang2009negative}
Zhang,~S.; Park,~Y.-S.; Li,~J.; Lu,~X.; Zhang,~W.; Zhang,~X. Negative
  refractive index in chiral metamaterials. \emph{Physical review letters}
  \textbf{2009}, \emph{102}, 023901\relax
\mciteBstWouldAddEndPuncttrue
\mciteSetBstMidEndSepPunct{\mcitedefaultmidpunct}
{\mcitedefaultendpunct}{\mcitedefaultseppunct}\relax
\EndOfBibitem
\bibitem[Lai \latin{et~al.}(2009)Lai, Ng, Chen, Han, Xiao, Zhang, and
  Chan]{lai2009illusion}
Lai,~Y.; Ng,~J.; Chen,~H.; Han,~D.; Xiao,~J.; Zhang,~Z.-Q.; Chan,~C.~T.
  Illusion optics: the optical transformation of an object into another object.
  \emph{Physical review letters} \textbf{2009}, \emph{102}, 253902\relax
\mciteBstWouldAddEndPuncttrue
\mciteSetBstMidEndSepPunct{\mcitedefaultmidpunct}
{\mcitedefaultendpunct}{\mcitedefaultseppunct}\relax
\EndOfBibitem
\bibitem[Mach-Batlle \latin{et~al.}(2018)Mach-Batlle, Parra, Laut, Del-Valle,
  Navau, and Sanchez]{mach2018magnetic}
Mach-Batlle,~R.; Parra,~A.; Laut,~S.; Del-Valle,~N.; Navau,~C.; Sanchez,~A.
  Magnetic illusion: transforming a magnetic object into another object by
  negative permeability. \emph{Physical Review Applied} \textbf{2018},
  \emph{9}, 034007\relax
\mciteBstWouldAddEndPuncttrue
\mciteSetBstMidEndSepPunct{\mcitedefaultmidpunct}
{\mcitedefaultendpunct}{\mcitedefaultseppunct}\relax
\EndOfBibitem
\bibitem[Barati \latin{et~al.}(2018)Barati, Fakheri, and
  Abdolali]{barati2018experimental}
Barati,~H.; Fakheri,~M.~H.; Abdolali,~A. Experimental demonstration of
  metamaterial-assisted antenna beam deflection through folded transformation
  optics. \emph{Journal of Optics} \textbf{2018}, \relax
\mciteBstWouldAddEndPunctfalse
\mciteSetBstMidEndSepPunct{\mcitedefaultmidpunct}
{}{\mcitedefaultseppunct}\relax
\EndOfBibitem
\bibitem[Pfeiffer \latin{et~al.}(2014)Pfeiffer, Emani, Shaltout, Boltasseva,
  Shalaev, and Grbic]{pfeiffer2014efficient}
Pfeiffer,~C.; Emani,~N.~K.; Shaltout,~A.~M.; Boltasseva,~A.; Shalaev,~V.~M.;
  Grbic,~A. Efficient light bending with isotropic metamaterial Huygens’
  surfaces. \emph{Nano letters} \textbf{2014}, \emph{14}, 2491--2497\relax
\mciteBstWouldAddEndPuncttrue
\mciteSetBstMidEndSepPunct{\mcitedefaultmidpunct}
{\mcitedefaultendpunct}{\mcitedefaultseppunct}\relax
\EndOfBibitem
\bibitem[Pollard \latin{et~al.}(2009)Pollard, Murphy, Hendren, Evans, Atkinson,
  Wurtz, Zayats, and Podolskiy]{pollard2009optical}
Pollard,~R.; Murphy,~A.; Hendren,~W.; Evans,~P.; Atkinson,~R.; Wurtz,~G.;
  Zayats,~A.; Podolskiy,~V.~A. Optical nonlocalities and additional waves in
  epsilon-near-zero metamaterials. \emph{Physical review letters}
  \textbf{2009}, \emph{102}, 127405\relax
\mciteBstWouldAddEndPuncttrue
\mciteSetBstMidEndSepPunct{\mcitedefaultmidpunct}
{\mcitedefaultendpunct}{\mcitedefaultseppunct}\relax
\EndOfBibitem
\bibitem[Alu \latin{et~al.}(2007)Alu, Silveirinha, Salandrino, and
  Engheta]{alu2007epsilon}
Alu,~A.; Silveirinha,~M.~G.; Salandrino,~A.; Engheta,~N. Epsilon-near-zero
  metamaterials and electromagnetic sources: Tailoring the radiation phase
  pattern. \emph{Physical review B} \textbf{2007}, \emph{75}, 155410\relax
\mciteBstWouldAddEndPuncttrue
\mciteSetBstMidEndSepPunct{\mcitedefaultmidpunct}
{\mcitedefaultendpunct}{\mcitedefaultseppunct}\relax
\EndOfBibitem
\bibitem[Liu \latin{et~al.}(2014)Liu, Zhang, Kenney, Su, Xu, Ouyang, Shi, Han,
  Zhang, and Zhang]{liu2014broadband}
Liu,~L.; Zhang,~X.; Kenney,~M.; Su,~X.; Xu,~N.; Ouyang,~C.; Shi,~Y.; Han,~J.;
  Zhang,~W.; Zhang,~S. Broadband metasurfaces with simultaneous control of
  phase and amplitude. \emph{Advanced Materials} \textbf{2014}, \emph{26},
  5031--5036\relax
\mciteBstWouldAddEndPuncttrue
\mciteSetBstMidEndSepPunct{\mcitedefaultmidpunct}
{\mcitedefaultendpunct}{\mcitedefaultseppunct}\relax
\EndOfBibitem
\bibitem[Rahmanzadeh \latin{et~al.}(2017)Rahmanzadeh, Rajabalipanah, and
  Abdolali]{rahmanzadeh2017analytical}
Rahmanzadeh,~M.; Rajabalipanah,~H.; Abdolali,~A. Analytical Investigation of
  Ultrabroadband Plasma--Graphene Radar Absorbing Structures. \emph{IEEE
  Transactions on Plasma Science} \textbf{2017}, \emph{45}, 945--954\relax
\mciteBstWouldAddEndPuncttrue
\mciteSetBstMidEndSepPunct{\mcitedefaultmidpunct}
{\mcitedefaultendpunct}{\mcitedefaultseppunct}\relax
\EndOfBibitem
\bibitem[Gao \latin{et~al.}(2015)Gao, Cheng, Yang, Ma, Zhao, Liu, Chen, He,
  Jiang, Ma, \latin{et~al.} others]{gao2015broadband}
Gao,~L.-H.; Cheng,~Q.; Yang,~J.; Ma,~S.-J.; Zhao,~J.; Liu,~S.; Chen,~H.-B.;
  He,~Q.; Jiang,~W.-X.; Ma,~H.-F., \latin{et~al.}  Broadband diffusion of
  terahertz waves by multi-bit coding metasurfaces. \emph{Light: Science \&
  Applications} \textbf{2015}, \emph{4}, e324\relax
\mciteBstWouldAddEndPuncttrue
\mciteSetBstMidEndSepPunct{\mcitedefaultmidpunct}
{\mcitedefaultendpunct}{\mcitedefaultseppunct}\relax
\EndOfBibitem
\bibitem[Rouhi \latin{et~al.}(2018)Rouhi, Rajabalipanah, and
  Abdolali]{rouhi2018real}
Rouhi,~K.; Rajabalipanah,~H.; Abdolali,~A. Real-Time and Broadband Terahertz
  Wave Scattering Manipulation via Polarization-Insensitive Conformal
  Graphene-Based Coding Metasurfaces. \emph{Annalen der Physik} \textbf{2018},
  \emph{530}, 1700310\relax
\mciteBstWouldAddEndPuncttrue
\mciteSetBstMidEndSepPunct{\mcitedefaultmidpunct}
{\mcitedefaultendpunct}{\mcitedefaultseppunct}\relax
\EndOfBibitem
\bibitem[Zhao and Al{\`u}(2011)Zhao, and Al{\`u}]{zhao2011manipulating}
Zhao,~Y.; Al{\`u},~A. Manipulating light polarization with ultrathin plasmonic
  metasurfaces. \emph{Physical Review B} \textbf{2011}, \emph{84}, 205428\relax
\mciteBstWouldAddEndPuncttrue
\mciteSetBstMidEndSepPunct{\mcitedefaultmidpunct}
{\mcitedefaultendpunct}{\mcitedefaultseppunct}\relax
\EndOfBibitem
\bibitem[Yin \latin{et~al.}(2015)Yin, Wan, Zhang, and Cui]{yin2015ultra}
Yin,~J.~Y.; Wan,~X.; Zhang,~Q.; Cui,~T.~J. Ultra wideband
  polarization-selective conversions of electromagnetic waves by metasurface
  under large-range incident angles. \emph{Scientific reports} \textbf{2015},
  \emph{5}, 12476\relax
\mciteBstWouldAddEndPuncttrue
\mciteSetBstMidEndSepPunct{\mcitedefaultmidpunct}
{\mcitedefaultendpunct}{\mcitedefaultseppunct}\relax
\EndOfBibitem
\bibitem[Liu \latin{et~al.}(2018)Liu, Zhu, Chen, Liang, and
  Zhu]{liu2018unidirectional}
Liu,~T.; Zhu,~X.; Chen,~F.; Liang,~S.; Zhu,~J. Unidirectional Wave Vector
  Manipulation in Two-Dimensional Space with an All Passive Acoustic
  Parity-Time-Symmetric Metamaterials Crystal. \emph{Physical review letters}
  \textbf{2018}, \emph{120}, 124502\relax
\mciteBstWouldAddEndPuncttrue
\mciteSetBstMidEndSepPunct{\mcitedefaultmidpunct}
{\mcitedefaultendpunct}{\mcitedefaultseppunct}\relax
\EndOfBibitem
\bibitem[Achouri \latin{et~al.}(2015)Achouri, Salem, and
  Caloz]{achouri2015general}
Achouri,~K.; Salem,~M.~A.; Caloz,~C. General metasurface synthesis based on
  susceptibility tensors. \emph{IEEE Transactions on Antennas and Propagation}
  \textbf{2015}, \emph{63}, 2977--2991\relax
\mciteBstWouldAddEndPuncttrue
\mciteSetBstMidEndSepPunct{\mcitedefaultmidpunct}
{\mcitedefaultendpunct}{\mcitedefaultseppunct}\relax
\EndOfBibitem
\bibitem[Momeni \latin{et~al.}(2018)Momeni, Rajabalipanah, Abdolali, and
  Achouri]{momeni2018generalized}
Momeni,~A.; Rajabalipanah,~H.; Abdolali,~A.; Achouri,~K. Generalized Optical
  Signal Processing Based on Multi-Operator Metasurfaces Synthesized by
  Susceptibility Tensors. \emph{arXiv preprint arXiv:1811.02618} \textbf{2018},
  \relax
\mciteBstWouldAddEndPunctfalse
\mciteSetBstMidEndSepPunct{\mcitedefaultmidpunct}
{}{\mcitedefaultseppunct}\relax
\EndOfBibitem
\bibitem[Cui \latin{et~al.}(2014)Cui, Qi, Wan, Zhao, and Cheng]{cui2014coding}
Cui,~T.~J.; Qi,~M.~Q.; Wan,~X.; Zhao,~J.; Cheng,~Q. Coding metamaterials,
  digital metamaterials and programmable metamaterials. \emph{Light: Science \&
  Applications} \textbf{2014}, \emph{3}, e218\relax
\mciteBstWouldAddEndPuncttrue
\mciteSetBstMidEndSepPunct{\mcitedefaultmidpunct}
{\mcitedefaultendpunct}{\mcitedefaultseppunct}\relax
\EndOfBibitem
\bibitem[Cui \latin{et~al.}(2016)Cui, Liu, and Li]{cui2016information}
Cui,~T.-J.; Liu,~S.; Li,~L.-L. Information entropy of coding metasurface.
  \emph{Light: Science \& Applications} \textbf{2016}, \emph{5}, e16172\relax
\mciteBstWouldAddEndPuncttrue
\mciteSetBstMidEndSepPunct{\mcitedefaultmidpunct}
{\mcitedefaultendpunct}{\mcitedefaultseppunct}\relax
\EndOfBibitem
\bibitem[Wan \latin{et~al.}(2016)Wan, Qi, Chen, and Cui]{wan2016field}
Wan,~X.; Qi,~M.~Q.; Chen,~T.~Y.; Cui,~T.~J. Field-programmable beam
  reconfiguring based on digitally-controlled coding metasurface.
  \emph{Scientific reports} \textbf{2016}, \emph{6}, 20663\relax
\mciteBstWouldAddEndPuncttrue
\mciteSetBstMidEndSepPunct{\mcitedefaultmidpunct}
{\mcitedefaultendpunct}{\mcitedefaultseppunct}\relax
\EndOfBibitem
\bibitem[Forouzmand and Mosallaei(2017)Forouzmand, and
  Mosallaei]{forouzmand2017real}
Forouzmand,~A.; Mosallaei,~H. Real-time controllable and multifunctional
  metasurfaces utilizing indium tin oxide materials: A phased array
  perspective. \emph{IEEE Transactions on Nanotechnology} \textbf{2017},
  \emph{16}, 296--306\relax
\mciteBstWouldAddEndPuncttrue
\mciteSetBstMidEndSepPunct{\mcitedefaultmidpunct}
{\mcitedefaultendpunct}{\mcitedefaultseppunct}\relax
\EndOfBibitem
\bibitem[D{\'\i}az-Rubio \latin{et~al.}(2017)D{\'\i}az-Rubio, Asadchy, Elsakka,
  and Tretyakov]{diaz2017generalized}
D{\'\i}az-Rubio,~A.; Asadchy,~V.~S.; Elsakka,~A.; Tretyakov,~S.~A. From the
  generalized reflection law to the realization of perfect anomalous
  reflectors. \emph{Science advances} \textbf{2017}, \emph{3}, e1602714\relax
\mciteBstWouldAddEndPuncttrue
\mciteSetBstMidEndSepPunct{\mcitedefaultmidpunct}
{\mcitedefaultendpunct}{\mcitedefaultseppunct}\relax
\EndOfBibitem
\bibitem[Chalabi \latin{et~al.}(2017)Chalabi, Ra'di, Sounas, and
  Al{\`u}]{chalabi2017efficient}
Chalabi,~H.; Ra'di,~Y.; Sounas,~D.; Al{\`u},~A. Efficient anomalous reflection
  through near-field interactions in metasurfaces. \emph{Physical Review B}
  \textbf{2017}, \emph{96}, 075432\relax
\mciteBstWouldAddEndPuncttrue
\mciteSetBstMidEndSepPunct{\mcitedefaultmidpunct}
{\mcitedefaultendpunct}{\mcitedefaultseppunct}\relax
\EndOfBibitem
\bibitem[Wong \latin{et~al.}(2014)Wong, Selvanayagam, and
  Eleftheriades]{wong2014design}
Wong,~J.~P.; Selvanayagam,~M.; Eleftheriades,~G.~V. Design of unit cells and
  demonstration of methods for synthesizing Huygens metasurfaces.
  \emph{Photonics and Nanostructures-Fundamentals and Applications}
  \textbf{2014}, \emph{12}, 360--375\relax
\mciteBstWouldAddEndPuncttrue
\mciteSetBstMidEndSepPunct{\mcitedefaultmidpunct}
{\mcitedefaultendpunct}{\mcitedefaultseppunct}\relax
\EndOfBibitem
\bibitem[Moccia \latin{et~al.}(2017)Moccia, Liu, Wu, Castaldi, Andreone, Cui,
  and Galdi]{moccia2017coding}
Moccia,~M.; Liu,~S.; Wu,~R.~Y.; Castaldi,~G.; Andreone,~A.; Cui,~T.~J.;
  Galdi,~V. Coding metasurfaces for diffuse scattering: scaling Laws, bounds,
  and suboptimal design. \emph{Advanced Optical Materials} \textbf{2017},
  \emph{5}, 1700455\relax
\mciteBstWouldAddEndPuncttrue
\mciteSetBstMidEndSepPunct{\mcitedefaultmidpunct}
{\mcitedefaultendpunct}{\mcitedefaultseppunct}\relax
\EndOfBibitem
\bibitem[Rajabalipanah \latin{et~al.}(2018)Rajabalipanah, Hemmati, Abdolali,
  and Amirhosseini]{rajabalipanah2018circular}
Rajabalipanah,~H.; Hemmati,~H.; Abdolali,~A.; Amirhosseini,~M.~K. Circular
  configuration of perforated dielectrics for ultra-broadband, wide-angle, and
  polarisation-insensitive monostatic/bistatic RCS reduction. \emph{IET
  Microwaves, Antennas \& Propagation} \textbf{2018}, \emph{12},
  1821--1827\relax
\mciteBstWouldAddEndPuncttrue
\mciteSetBstMidEndSepPunct{\mcitedefaultmidpunct}
{\mcitedefaultendpunct}{\mcitedefaultseppunct}\relax
\EndOfBibitem
\bibitem[Momeni \latin{et~al.}(2018)Momeni, Rouhi, Rajabalipanah, and
  Abdolali]{momeni2018information}
Momeni,~A.; Rouhi,~K.; Rajabalipanah,~H.; Abdolali,~A. An Information
  Theory-Inspired Strategy for Design of Re-programmable Encrypted
  Graphene-based Coding Metasurfaces at Terahertz Frequencies. \emph{Scientific
  reports} \textbf{2018}, \emph{8}, 6200\relax
\mciteBstWouldAddEndPuncttrue
\mciteSetBstMidEndSepPunct{\mcitedefaultmidpunct}
{\mcitedefaultendpunct}{\mcitedefaultseppunct}\relax
\EndOfBibitem
\bibitem[Ma \latin{et~al.}(2016)Ma, Liu, Luan, and Cui]{ma2016multi}
Ma,~H.~F.; Liu,~Y.~Q.; Luan,~K.; Cui,~T.~J. Multi-beam reflections with
  flexible control of polarizations by using anisotropic metasurfaces.
  \emph{Scientific reports} \textbf{2016}, \emph{6}, 39390\relax
\mciteBstWouldAddEndPuncttrue
\mciteSetBstMidEndSepPunct{\mcitedefaultmidpunct}
{\mcitedefaultendpunct}{\mcitedefaultseppunct}\relax
\EndOfBibitem
\bibitem[Wu \latin{et~al.}(2018)Wu, Shi, Liu, Wu, and Cui]{wu2018addition}
Wu,~R.~Y.; Shi,~C.~B.; Liu,~S.; Wu,~W.; Cui,~T.~J. Addition Theorem for Digital
  Coding Metamaterials. \emph{Advanced Optical Materials} \textbf{2018},
  \emph{6}, 1701236\relax
\mciteBstWouldAddEndPuncttrue
\mciteSetBstMidEndSepPunct{\mcitedefaultmidpunct}
{\mcitedefaultendpunct}{\mcitedefaultseppunct}\relax
\EndOfBibitem
\bibitem[Zhang \latin{et~al.}(2018)Zhang, Deng, Yang, Jiang, Xu, and
  Li]{zhang2018metasurface}
Zhang,~X.; Deng,~R.; Yang,~F.; Jiang,~C.; Xu,~S.; Li,~M. Metasurface-Based
  Ultrathin Beam Splitter with Variable Split Angle and Power Distribution.
  \emph{ACS Photonics} \textbf{2018}, \emph{5}, 2997--3002\relax
\mciteBstWouldAddEndPuncttrue
\mciteSetBstMidEndSepPunct{\mcitedefaultmidpunct}
{\mcitedefaultendpunct}{\mcitedefaultseppunct}\relax
\EndOfBibitem
\bibitem[Nayeri \latin{et~al.}(2013)Nayeri, Yang, and
  Elsherbeni]{nayeri2013design}
Nayeri,~P.; Yang,~F.; Elsherbeni,~A.~Z. Design of single-feed reflectarray
  antennas with asymmetric multiple beams using the particle swarm optimization
  method. \emph{IEEE Trans. Antennas Propag} \textbf{2013}, \emph{61},
  4598--4605\relax
\mciteBstWouldAddEndPuncttrue
\mciteSetBstMidEndSepPunct{\mcitedefaultmidpunct}
{\mcitedefaultendpunct}{\mcitedefaultseppunct}\relax
\EndOfBibitem
\bibitem[G{\'o}mez \latin{et~al.}(2014)G{\'o}mez, Tayebi, and
  C{\'a}tedra]{gomez2014optimization}
G{\'o}mez,~J.; Tayebi,~A.; C{\'a}tedra,~F. Optimization Approach to Design
  Single Feed Symmetric and Asymmetric Multibeam Reflectarrays. \emph{Frequenz}
  \textbf{2014}, \emph{68}, 531--536\relax
\mciteBstWouldAddEndPuncttrue
\mciteSetBstMidEndSepPunct{\mcitedefaultmidpunct}
{\mcitedefaultendpunct}{\mcitedefaultseppunct}\relax
\EndOfBibitem
\bibitem[Zhao \latin{et~al.}(2018)Zhao, Yang, Dai, Cheng, Li, Qi, Ke, Bai, Liu,
  Jin, \latin{et~al.} others]{zhao2018programmable}
Zhao,~J.; Yang,~X.; Dai,~J.~Y.; Cheng,~Q.; Li,~X.; Qi,~N.~H.; Ke,~J.~C.;
  Bai,~G.~D.; Liu,~S.; Jin,~S., \latin{et~al.}  Programmable time-domain
  digital coding metasurface for nonlinear harmonic manipulation and new
  wireless communication systems. \emph{National Science Review} \textbf{2018},
  \relax
\mciteBstWouldAddEndPunctfalse
\mciteSetBstMidEndSepPunct{\mcitedefaultmidpunct}
{}{\mcitedefaultseppunct}\relax
\EndOfBibitem
\bibitem[Tang \latin{et~al.}(2018)Tang, Li, Dai, Jin, Zeng, Cheng, and
  Cui]{tang2018wireless}
Tang,~W.; Li,~X.; Dai,~J.~Y.; Jin,~S.; Zeng,~Y.; Cheng,~Q.; Cui,~T.~J. Wireless
  Communications with Programmable Metasurface: Transceiver Design and
  Experimental Results. \emph{arXiv preprint arXiv:1811.08119} \textbf{2018},
  \relax
\mciteBstWouldAddEndPunctfalse
\mciteSetBstMidEndSepPunct{\mcitedefaultmidpunct}
{}{\mcitedefaultseppunct}\relax
\EndOfBibitem
\bibitem[Zhang \latin{et~al.}(2018)Zhang, Tang, Jiang, Bai, Tang, Bai, Qiu, and
  Cui]{zhang2018digital}
Zhang,~X.~G.; Tang,~W.~X.; Jiang,~W.~X.; Bai,~G.~D.; Tang,~J.; Bai,~L.;
  Qiu,~C.-W.; Cui,~T.~J. Digital Metasurfaces: Light-Controllable Digital
  Coding Metasurfaces (Adv. Sci. 11/2018). \emph{Advanced Science}
  \textbf{2018}, \emph{5}, 1870068\relax
\mciteBstWouldAddEndPuncttrue
\mciteSetBstMidEndSepPunct{\mcitedefaultmidpunct}
{\mcitedefaultendpunct}{\mcitedefaultseppunct}\relax
\EndOfBibitem
\bibitem[Bao \latin{et~al.}(2018)Bao, Ma, Bai, Jing, Wu, Fu, Yang, Wu, and
  Cui]{bao2018design}
Bao,~L.; Ma,~Q.; Bai,~G.~D.; Jing,~H.~B.; Wu,~R.~Y.; Fu,~X.; Yang,~C.; Wu,~J.;
  Cui,~T.~J. Design of digital coding metasurfaces with independent controls of
  phase and amplitude responses. \emph{Applied Physics Letters} \textbf{2018},
  \emph{113}, 063502\relax
\mciteBstWouldAddEndPuncttrue
\mciteSetBstMidEndSepPunct{\mcitedefaultmidpunct}
{\mcitedefaultendpunct}{\mcitedefaultseppunct}\relax
\EndOfBibitem
\bibitem[Zhang \latin{et~al.}(2017)Zhang, Liu, Li, and Cui]{zhang2017spin}
Zhang,~L.; Liu,~S.; Li,~L.; Cui,~T.~J. Spin-Controlled Multiple Pencil Beams
  and Vortex Beams with Different Polarizations Generated by Pancharatnam-Berry
  Coding Metasurfaces. \emph{ACS applied materials \& interfaces}
  \textbf{2017}, \emph{9}, 36447--36455\relax
\mciteBstWouldAddEndPuncttrue
\mciteSetBstMidEndSepPunct{\mcitedefaultmidpunct}
{\mcitedefaultendpunct}{\mcitedefaultseppunct}\relax
\EndOfBibitem
\bibitem[Liu \latin{et~al.}(2016)Liu, Noor, Du, Zhang, Xu, Luan, Wang, Tian,
  Tang, Han, \latin{et~al.} others]{liu2016anomalous}
Liu,~S.; Noor,~A.; Du,~L.~L.; Zhang,~L.; Xu,~Q.; Luan,~K.; Wang,~T.~Q.;
  Tian,~Z.; Tang,~W.~X.; Han,~J.~G., \latin{et~al.}  Anomalous refraction and
  nondiffractive Bessel-beam generation of terahertz waves through
  transmission-type coding metasurfaces. \emph{ACS photonics} \textbf{2016},
  \emph{3}, 1968--1977\relax
\mciteBstWouldAddEndPuncttrue
\mciteSetBstMidEndSepPunct{\mcitedefaultmidpunct}
{\mcitedefaultendpunct}{\mcitedefaultseppunct}\relax
\EndOfBibitem
\bibitem[Yu \latin{et~al.}(2011)Yu, Genevet, Kats, Aieta, Tetienne, Capasso,
  and Gaburro]{yu2011light}
Yu,~N.; Genevet,~P.; Kats,~M.~A.; Aieta,~F.; Tetienne,~J.-P.; Capasso,~F.;
  Gaburro,~Z. Light propagation with phase discontinuities: generalized laws of
  reflection and refraction. \emph{science} \textbf{2011}, 1210713\relax
\mciteBstWouldAddEndPuncttrue
\mciteSetBstMidEndSepPunct{\mcitedefaultmidpunct}
{\mcitedefaultendpunct}{\mcitedefaultseppunct}\relax
\EndOfBibitem
\bibitem[Ding \latin{et~al.}(2017)Ding, Pors, and
  Bozhevolnyi]{ding2017gradient}
Ding,~F.; Pors,~A.; Bozhevolnyi,~S.~I. Gradient metasurfaces: a review of
  fundamentals and applications. \emph{Reports on Progress in Physics}
  \textbf{2017}, \emph{81}, 026401\relax
\mciteBstWouldAddEndPuncttrue
\mciteSetBstMidEndSepPunct{\mcitedefaultmidpunct}
{\mcitedefaultendpunct}{\mcitedefaultseppunct}\relax
\EndOfBibitem
\bibitem[Jiu-sheng \latin{et~al.}(2017)Jiu-sheng, Ze-jiang, and
  Jian-quan]{jiu2017flexible}
Jiu-sheng,~L.; Ze-jiang,~Z.; Jian-quan,~Y. Flexible manipulation of terahertz
  wave reflection using polarization insensitive coding metasurfaces.
  \emph{Optics express} \textbf{2017}, \emph{25}, 29983--29992\relax
\mciteBstWouldAddEndPuncttrue
\mciteSetBstMidEndSepPunct{\mcitedefaultmidpunct}
{\mcitedefaultendpunct}{\mcitedefaultseppunct}\relax
\EndOfBibitem
\bibitem[Liang \latin{et~al.}(2016)Liang, Wei, Yan, Wei, Liang, Han, Ding,
  Zhang, and Yao]{liang2016broadband}
Liang,~L.; Wei,~M.; Yan,~X.; Wei,~D.; Liang,~D.; Han,~J.; Ding,~X.; Zhang,~G.;
  Yao,~J. Broadband and wide-angle RCS reduction using a 2-bit coding ultrathin
  metasurface at terahertz frequencies. \emph{Scientific reports}
  \textbf{2016}, \emph{6}, 39252\relax
\mciteBstWouldAddEndPuncttrue
\mciteSetBstMidEndSepPunct{\mcitedefaultmidpunct}
{\mcitedefaultendpunct}{\mcitedefaultseppunct}\relax
\EndOfBibitem
\bibitem[Liu \latin{et~al.}(2016)Liu, Cui, Zhang, Xu, Wang, Wan, Gu, Tang,
  Qing~Qi, Han, \latin{et~al.} others]{liu2016convolution}
Liu,~S.; Cui,~T.~J.; Zhang,~L.; Xu,~Q.; Wang,~Q.; Wan,~X.; Gu,~J.~Q.;
  Tang,~W.~X.; Qing~Qi,~M.; Han,~J.~G., \latin{et~al.}  Convolution operations
  on coding metasurface to reach flexible and continuous controls of terahertz
  beams. \emph{Advanced Science} \textbf{2016}, \emph{3}, 1600156\relax
\mciteBstWouldAddEndPuncttrue
\mciteSetBstMidEndSepPunct{\mcitedefaultmidpunct}
{\mcitedefaultendpunct}{\mcitedefaultseppunct}\relax
\EndOfBibitem
\bibitem[Nayeri \latin{et~al.}(2012)Nayeri, Yang, and
  Elsherbeni]{nayeri2012design}
Nayeri,~P.; Yang,~F.; Elsherbeni,~A.~Z. Design and experiment of a single-feed
  quad-beam reflectarray antenna. \emph{IEEE Transactions on Antennas and
  Propagation} \textbf{2012}, \emph{60}, 1166--1171\relax
\mciteBstWouldAddEndPuncttrue
\mciteSetBstMidEndSepPunct{\mcitedefaultmidpunct}
{\mcitedefaultendpunct}{\mcitedefaultseppunct}\relax
\EndOfBibitem
\bibitem[Huang \latin{et~al.}(2018)Huang, Zhang, and
  Zentgraf]{huang2018metasurface}
Huang,~L.; Zhang,~S.; Zentgraf,~T. Metasurface holography: from fundamentals to
  applications. \emph{Nanophotonics} \textbf{2018}, \emph{7}, 1169--1190\relax
\mciteBstWouldAddEndPuncttrue
\mciteSetBstMidEndSepPunct{\mcitedefaultmidpunct}
{\mcitedefaultendpunct}{\mcitedefaultseppunct}\relax
\EndOfBibitem
\bibitem[Liu and Cui(2017)Liu, and Cui]{liu2017concepts}
Liu,~S.; Cui,~T.~J. Concepts, working principles, and applications of coding
  and programmable metamaterials. \emph{Advanced Optical Materials}
  \textbf{2017}, \emph{5}, 1700624\relax
\mciteBstWouldAddEndPuncttrue
\mciteSetBstMidEndSepPunct{\mcitedefaultmidpunct}
{\mcitedefaultendpunct}{\mcitedefaultseppunct}\relax
\EndOfBibitem
\bibitem[Yan \latin{et~al.}(2015)Yan, Liang, Yang, Liu, Ding, Xu, Zhang, Cui,
  and Yao]{yan2015broadband}
Yan,~X.; Liang,~L.; Yang,~J.; Liu,~W.; Ding,~X.; Xu,~D.; Zhang,~Y.; Cui,~T.;
  Yao,~J. Broadband, wide-angle, low-scattering terahertz wave by a flexible
  2-bit coding metasurface. \emph{Optics express} \textbf{2015}, \emph{23},
  29128--29137\relax
\mciteBstWouldAddEndPuncttrue
\mciteSetBstMidEndSepPunct{\mcitedefaultmidpunct}
{\mcitedefaultendpunct}{\mcitedefaultseppunct}\relax
\EndOfBibitem
\bibitem[Ding \latin{et~al.}(2016)Ding, Xu, Ren, Lin, Zhang, and
  Zhang]{ding2016dual}
Ding,~J.; Xu,~N.; Ren,~H.; Lin,~Y.; Zhang,~W.; Zhang,~H. Dual-wavelength
  terahertz metasurfaces with independent phase and amplitude control at each
  wavelength. \emph{Scientific reports} \textbf{2016}, \emph{6}, 34020\relax
\mciteBstWouldAddEndPuncttrue
\mciteSetBstMidEndSepPunct{\mcitedefaultmidpunct}
{\mcitedefaultendpunct}{\mcitedefaultseppunct}\relax
\EndOfBibitem
\bibitem[Farmahini-Farahani \latin{et~al.}(2013)Farmahini-Farahani, Cheng, and
  Mosallaei]{farmahini2013metasurfaces}
Farmahini-Farahani,~M.; Cheng,~J.; Mosallaei,~H. Metasurfaces nanoantennas for
  light processing. \emph{JOSA B} \textbf{2013}, \emph{30}, 2365--2370\relax
\mciteBstWouldAddEndPuncttrue
\mciteSetBstMidEndSepPunct{\mcitedefaultmidpunct}
{\mcitedefaultendpunct}{\mcitedefaultseppunct}\relax
\EndOfBibitem
\bibitem[Wan \latin{et~al.}(2016)Wan, Jia, Cui, and Zhao]{wan2016independent}
Wan,~X.; Jia,~S.~L.; Cui,~T.~J.; Zhao,~Y.~J. Independent modulations of the
  transmission amplitudes and phases by using Huygens metasurfaces.
  \emph{Scientific reports} \textbf{2016}, \emph{6}, 25639\relax
\mciteBstWouldAddEndPuncttrue
\mciteSetBstMidEndSepPunct{\mcitedefaultmidpunct}
{\mcitedefaultendpunct}{\mcitedefaultseppunct}\relax
\EndOfBibitem
\bibitem[Lee \latin{et~al.}(2018)Lee, Yoon, Lee, Yun, Cho, Lee, Kim, Rho, and
  Lee]{lee2018complete}
Lee,~G.-Y.; Yoon,~G.; Lee,~S.-Y.; Yun,~H.; Cho,~J.; Lee,~K.; Kim,~H.; Rho,~J.;
  Lee,~B. Complete amplitude and phase control of light using broadband
  holographic metasurfaces. \emph{Nanoscale} \textbf{2018}, \emph{10},
  4237--4245\relax
\mciteBstWouldAddEndPuncttrue
\mciteSetBstMidEndSepPunct{\mcitedefaultmidpunct}
{\mcitedefaultendpunct}{\mcitedefaultseppunct}\relax
\EndOfBibitem
\bibitem[Zhu \latin{et~al.}(2017)Zhu, Zhou, Lou, Ye, Qiu, Ruan, and
  Fan]{zhu2017plasmonic}
Zhu,~T.; Zhou,~Y.; Lou,~Y.; Ye,~H.; Qiu,~M.; Ruan,~Z.; Fan,~S. Plasmonic
  computing of spatial differentiation. \emph{Nature communications}
  \textbf{2017}, \emph{8}, 15391\relax
\mciteBstWouldAddEndPuncttrue
\mciteSetBstMidEndSepPunct{\mcitedefaultmidpunct}
{\mcitedefaultendpunct}{\mcitedefaultseppunct}\relax
\EndOfBibitem
\end{mcitethebibliography}

\end{document}